\documentclass[preprint,5p,twocolumn,number,sort&compress]{elsarticle}{\tiny}

\usepackage[nodisplayskipstretch]{setspace}

\usepackage{epsfig,color,amsmath} 

\usepackage{amsthm} 
\usepackage{amsmath}    
\usepackage{bm}
\usepackage{epstopdf}
\usepackage{amssymb}
\usepackage{url}
\usepackage{enumitem} 
\usepackage{multirow}
\usepackage{hhline}
\usepackage{booktabs}
\usepackage{mathtools}
\usepackage{makecell}
\usepackage[linesnumbered,boxed,commentsnumbered,ruled,vlined,longend]{algorithm2e}
\usepackage{comment}

\DeclareMathOperator*{\minimize}{minimize}

\DeclareMathOperator*{\subjectto}{subject\ to}

\DeclareMathOperator*{\argmin}{arg\ min}
\makeatother
\DeclareMathAlphabet\mathbfcal{OMS}{cmsy}{b}{n}

\newtheorem{theorem}{Theorem}
\newtheorem{mydef}{Definition}

\newtheorem{asmp}{Assumption}
\newtheorem{mycor}{Corollary}



\usepackage{stackengine}
\newcommand\barbelow[1]{\stackunder[1.2pt]{$#1$}{\rule{.8ex}{.075ex}}}

\newcommand{\mat}[1]{\boldsymbol{#1}}

\newcommand{\bmat}[1]{\begin{bmatrix} #1 \end{bmatrix}}

\providecommand{\mA}{\ensuremath{\mat{A}}}
\providecommand{\mB}{\ensuremath{\mat{B}}}
\providecommand{\mC}{\ensuremath{\mat{C}}}

\providecommand{\mG}{\ensuremath{\mat{G}}}

\providecommand{\mI}{\ensuremath{\mat{I}}}

\providecommand{\mK}{\ensuremath{\mat{K}}}
\providecommand{\mL}{\ensuremath{\mat{L}}}

\providecommand{\mO}{\ensuremath{\mat{O}}}
\providecommand{\mP}{\ensuremath{\mat{P}}}

\providecommand{\mY}{\ensuremath{\mat{Y}}}
\providecommand{\mZ}{\ensuremath{\mat{Z}}}

\newcommand{\m}{\boldsymbol}
\allowdisplaybreaks[4]
\newcommand{\mbb}[1]{\mathbb{#1}}

\usepackage[framemethod=TikZ]{mdframed}
\mdfdefinestyle{MyFrame}{%
	linecolor=black,
	outerlinewidth=1.25pt,
	roundcorner=1.25pt,
	innerrightmargin=5pt,
	innerleftmargin=5pt,}
	
\usepackage[noabbrev]{cleveref}

\usepackage{mathtools}

\DeclarePairedDelimiter\abs{\lvert}{\rvert}%
\DeclarePairedDelimiter\norm{\lVert}{\rVert}%

\makeatletter
\let\oldabs\abs
\def\abs{\@ifstar{\oldabs}{\oldabs*}}
\let\oldnorm\norm
\def\norm{\@ifstar{\oldnorm}{\oldnorm*}}
\makeatother

\usepackage[english]{babel}
\usepackage[utf8]{inputenc}
\usepackage[super]{nth}

\usepackage{graphicx}
\usepackage{float}
\usepackage[caption = false]{subfig}

\usepackage{array}
\usepackage{threeparttable}

\usepackage[english]{babel}
\usepackage[utf8]{inputenc}
\usepackage[super]{nth}

\RequirePackage{filecontents}

\def\newqed{{\null\nobreak\hfill\color{black}\ensuremath{\blacksquare}}}

\SetKwRepeat{Do}{do}{while}%

\usepackage{nccmath}
\usepackage{lettrine}

\usepackage{enumitem}
\setitemize{itemsep=-1pt}

\journal{Automatica}

\begin{document}
	
\begin{frontmatter}
	
\let\WriteBookmarks\relax
\def\floatpagepagefraction{1}
\def\textpagefraction{.001}

\setlength{\abovedisplayskip}{3pt}
\setlength{\belowdisplayskip}{3pt}

\newdimen\origiwspc%
\newdimen\origiwstr%
\origiwspc=\fontdimen2\font
\origiwstr=\fontdimen3\font

\fontdimen2\font=0.63ex


\title{Towards Understanding Sensor and Control Nodes Selection in Nonlinear Dynamic Systems: Lyapunov Theory Meets Branch-and-Bound}

\author[1]{Sebastian A. Nugroho}

\ead{sebastian.nugroho@my.utsa.edu}

\author[1]{Ahmad F. Taha\corref{cor1}}

\ead{ahmad.taha@utsa.edu}

\cortext[cor1]{Corresponding author. This material is based upon work supported by the National Science Foundation under Grants 1728629, 1917164, 2151571, and 2152450.}

\address[1]{Department of Electrical and Computer Engineering, The University of Texas at San Antonio, 1 UTSA Circle, San Antonio, TX 78249}

\begin{abstract}
{\textcolor{black}{Sensor and actuator selection problems (SASPs) are some of the core problems in dynamic systems design and control. These problems correspond to determining the optimal selection of sensors (measurements) or actuators (control nodes) such that certain estimation/control objectives can be achieved. While the literature on SASPs are indeed inveterate, the vast majority of the work focuses on linear(ized) representation of the network dynamics, resulting in the placements of sensors or actuators (SAs) that are valid for confined operating regions.} As an alternative, herein we propose a new general framework for addressing SASPs in nonlinear dynamic systems (NDSs), assuming that the inputs and outputs are linearly coupled with the nonlinear dynamics. This is investigated through \textit{(i)} classifying and parameterizing the NDSs into various nonlinear function sets, \textit{(ii)} utilizing rich Lyapunov theoretic formulations, and \textit{(iii)} designing a new customized branch-and-bound (BnB) algorithm that exploits problem structure of the SASPs. The newly designed BnB routines are computationally more attractive than the standard one and also directly applicable to solve SASPs for linear systems. In contrast with contemporary approaches from the literature, our approach is suitable for finding the optimal SAs combination for stable/unstable NDSs that ensures stabilization of estimation error and closed-loop dynamics through a simple linear feedback control policy.}
\end{abstract}

%


\end{frontmatter}

\setlength{\abovedisplayskip}{3pt}
\setlength{\belowdisplayskip}{3pt}

\newdimen\origiwspc%
\newdimen\origiwstr%
\origiwspc=\fontdimen2\font
\origiwstr=\fontdimen3\font

\fontdimen2\font=0.63ex


\section{Introduction and Paper Contributions}
\vspace{-0.20cm}
{Sensor and actuator selection problems (SASPs)---which fall into \textit{(i)} finding optimal geographic placements or time-varying selection of sensors or actuators (SAs) (but not necessarily simultaneously), while \textit{(ii)} optimizing state estimation or control metrics---have become prevalent research topics in numerous science and engineering fields.} For instance, in power networks application, sensor selection corresponds to the placement of phasor measurement units for the purpose of power systems monitoring \citep{qi2015optimal} while actuator selection corresponds to the placement of energy storage devices to ensure system controllability \citep{Pequito2013}. Both problems are crucial to achieve a more economical and robust power systems operations while ensuring that certain minimal requirements, such as allowable tolerance on dynamic estimation error and voltage/frequency deviation, are met.

%

{Extensive studies have been carried out in the literature to address SASPs for particularly \textit{linear} dynamic systems---see \citep{SUMMERS20143784,Dhingra2014,Taha2018,Nugroho2019124} for some notable references.}
Unlike SASPs in linear(ized) systems which yield feasible or optimal SAs configurations for specific operating points, SASPs for \textit{nonlinear} dynamic systems (NDSs) offer SA selections that are applicable for a much larger operating regions or even globally---this is demonstrated in our prior work through a simple example \citep{Nugroho2019sap}. 
Indeed, SASPs for NDSs have not received comparable attention in the literature as the SASPs for linear systems, until recently.
{In \citep{qi2015optimal}, the concept of \textit{empirical} Gramian to quantify the observability of NDSs is considered jointly with the sensor selection problem (SSP) to determine the combination of sensors that maximizes the logarithmic determinant of the empirical Gramian matrix. The joint problem of sensor selection and state observation is investigated in \citep{Haber2017}. In particular, a new method for reconstructing the initial states of NDSs while simultaneously selecting sensors for a given observation window is presented. A different approach is pursued in \citep{Bopardikar2019} where the authors introduce a randomized algorithm for dealing with SSP and develop theoretical bounds for eigenvalue and condition number of observability Gramian. Recently, a method for selecting the control nodes and designing control actions for NDSs, which is based on an open-loop predictive control framework, is proposed in \cite{Haber2021}.} In spite of these efforts, their potential applicability to address SASPs in unstable NDSs along with their ability to incorporate SAs selection with some estimation/control-theoretic metrics, unfortunately, remains unclear. 

The objective of this paper is to find the minimal SAs combination for stable/unstable NDSs that ensures stabilization for estimation error (via minimal sensor selection) and closed loop system through feedback control (via minimal actuator selection). The proposed algorithms developed herein are based on the following observations. First, numerous observer and controller designs for NDSs have been developed in the past two decades. By using Lyapunov stability theory and given that the nonlinearities in NDSs satisfy some function sets such as \textit{bounded Jacobian} \citep{Phanomchoeng2010b}, \textit{Lipschitz continuous} \citep{Phanomchoeng2010}, \textit{one-sided Lipschitz} \citep{Abbaszadeh2010}, \textit{quadratically inner-bounded} \citep{Abbaszadeh2010}, and \textit{quadratically bounded} \citep{guo2017decentralized}, the procedure to compute stabilizing observer/controller gain matrix can be posed as convex semidefinite programmings (SDPs).  
{Secondly, physical states in many practical NDS models are almost always bounded. 
Hence, it can be shown or safely assumed that the corresponding nonlinearities satisfy the properties of the aforementioned function sets in confined state-space regions---regions that are much larger than operating regions of linearization points.}

Without loss of generality, this paper focuses on SASPs for Lipschitz NDSs in which the vector-valued nonlinearity in the system dynamics satisfies the Lipschitz continuity assumption. The proposed method can be easily extended for other types of non-Lipschitz nonlinearities.
{It is assumed throughout the paper that the inputs and outputs of the NDSs form linear relations with the nonlinear dynamics.
It is also worth noting that we do \textit{not} address the simultaneous selection of SAs. That is, we only consider the minimal selection of sensors with observer design or actuators with controller design.
A preliminary version of this work appeared in \citep{Nugroho2019CDC}, where the SSP for Lipschitz NDSs is studied and solved via a general purpose solver \citep{Lofberg2004}. The key contributions of this paper are summarized as follows:}
\vspace{-0.2cm}
\begin{itemize}[leftmargin=*]
	\setlength\itemsep{-0.3em}
	\item We present a novel generalized framework to address SASPs in NDSs by:  \textit{(i)} parameterizing NDSs' nonlinearities based on the corresponding function sets, \textit{(ii)} formulating the SASPs using a plethora of observer and/or controller designs, and \textit{(iii)} reformulating the resulting SASPs into convex mixed-integer SDPs (MISDPs).
	\item Particularly for Lipschitz NDSs with Lipschitz constant $\gamma_l$, we formalize NDSs observability and controllability (coined as \textit{$\gamma_l$-observability} and \textit{$\gamma_l$-controllability}, respectively) as SDP feasibility problems and show that it is crucial to obtain the smallest Lipschitz constant in order to obtain less number of SAs in SASPs. {In particular, we theoretically investigate the relationship between the Lipschitz constant of the system and the number of activated SAs}. 
	\item {A new customized branch-and-bound (BnB) algorithm for  MISDPs to solve SASPs is proposed. This BnB algorithm, referred to as \textit{structure-exploiting} BnB (SE-BnB), utilizes heuristics and exploits problem structure to efficiently find optimal and suboptimal solutions for SASPs, owing to the reduced number of constraints and/or variables of the corresponding SDP problems when finding upper an lower bounds on each node of the BnB tree.}
	\item {We showcase the computational advantages of the proposed BnB routines for solving SSP in comparison with our implementation of the standard BnB algorithm and the $\ell_1$-norm relaxation technique \citep{Candes2008}.}
\end{itemize}   
\vspace{-0.2cm}
The distinction of our approach to address SASPs for NDSs in comparison with the other contemporary methods includes \textit{(a)} its ability to find optimal placement of SAs with stabilizing estimation/control gain matrix, and \textit{(b)} its flexibility to incorporate estimation/control metrics for continuous-time/discrete-time NDSs having different classes of nonlinearity. The paper's organization and notation are as follows. 
{Section \ref{sec:model_problem} formalizes the NDSs model. Section \ref{sec:ssp} investigates the relation between Lipschitz constant and $\gamma_l$-observability while Section \ref{sec:obs_ssp_sensor} constructs the SSP for Lipschitz NDSs. Next, Section \ref{sec:sol_appr} focuses on transforming SSP into a convex MISDP form and provides a detailed description for the SE-BnB algorithm. In Section \ref{sec:misc}, some discussions related to actuator selection problem, solving SASPs
beyond Lipschitz NDSs, and robust SSP are provided. Finally, Section \ref{sec:num_test} presents some numerical results and Section \ref{sec:conclusion} concludes the paper.}

\noindent {\textbf{Notation.}} \hspace{0.5cm} The notation $\m 1$ represents a matrix of appropriate dimension with elements of 1.
The notations $\mathbb{R}^n$ and $\mathbb{R}^{p\times q}$ denote the sets of row vectors with $n$ elements and matrices with size $p$-by-$q$ with elements in $\mathbb{R}$. 
Given $\m A$ as a matrix, its $i$-th and $j$-th element is denoted by $A_{ij}$. The operators $\mathrm{Blkdiag}(\cdot)$ constructs a block diagonal matrix, $\mathrm{Diag}(\cdot)$ constructs a diagonal matrix from a vector, $\mathrm{vec}(\cdot)$ constructs a vector by stacking each column of a matrix, $\otimes$ denotes the Kronecker product, and $\odot$ denotes the Hadamard product. 
The symbol $*$ is used to represent symmetric entries in symmetric matrices.
The set $\mathbb{I}(n)$ is defined as $\mathbb{I}(n) := \{i\in\mathbb{N}\,|\, 1\leq i \leq n\}$, which is usually used to represent the set of indices. 

\setlength{\textfloatsep}{10pt}
\begin{table*}
	\vspace{-0.05cm}
	\begin{center}
		{	\centering \scriptsize
			\caption{Dynamic models of some nonlinear dynamic systems in electric power systems, traffic, combustion networks, \ldots.}\label{tab:nonlinear_models_cps}
			\vspace{-0.0cm}
			\renewcommand{\arraystretch}{1}
			\begin{tabular}{c|c|c}
				\toprule \midrule
				\textbf{Type of Systems}	& \textbf{Nonlinear Dynamics Model} & \textbf{Description}   \\ \midrule
				\makecell{\textit{Electric power}\\ \textit{grids}\citep{bergen2000power}}	& \makecell{$\dot{\delta_i}=\omega_i-\omega_0\qquad \qquad\qquad\qquad\qquad\qquad\qquad\qquad\qquad\qquad\qquad\qquad\qquad\qquad\qquad\;$\\  $\hspace{-.25cm}\dot{\omega}_i=\frac{\omega_0}{2H_i}\bigg(\hspace{-0.05cm}P_{\textrm{m}i}-\sum\limits_{j=1 }^{N} \hspace{-0.05cm}E_iE_j\hspace{-0.05cm}\left(\bar{G}_{ij}\cos(\delta_i-\delta_j)+\bar{B}_{ij}\sin(\delta_i-\delta_j)\right)-\frac{K_{\textrm{D}i}}{\omega_0}(\omega_i-\omega_0)\hspace{-0.05cm}\bigg)$} & \makecell{\nth{2}-order swing equation \\ $\m{x_i} := [\delta_i\;\;\omega_i]^{\top}$ \\
					$\delta_i$: rotor angle,
					$\omega_i$: rotor speed} \\ \midrule
				\textit{Highway traffic} \citep{nugroho2018journal}	& $\dot{\rho}_i = \frac{v_f}{l}\left(\rho_{i-1}-\rho_i+\hat{\rho}_j-\alpha(k)\check{\rho}_k\right) -\delta \left(\rho_{i-1}^2-\rho_i^2+\hat{\rho}_j^2-\alpha(k)\check{\rho}_k^2\right)$ &  \makecell{Free-flow condition \\ $\m{x_i} := [\rho_i]$, $\rho_i$: traffic density} \\ \midrule
				\makecell{\textit{Combustion} \\  \textit{networks}\citep{Perini2012}}	&  $\dot{\chi}_i = \sum\limits_{j = 1}^{n_r}(\beta_{ji}-\alpha_{ji})\Bigg(\hspace{-0.05cm}d^{(f)}_j\prod\limits^n_{k=1}\chi_k^{\alpha_{jk}}-d^{(b)}_j\prod\limits^n_{k=1}\chi_k^{\beta_{jk}}\hspace{-0.05cm}\Bigg)$ & \makecell{$\m{x_i} := [\chi_i]$ \\ $\chi_i$: chemical concentration} \\ \midrule
				\makecell{\textit{Oscillator} \\ \textit{synchronization} \citep{Kuramoto1975ebm}}	& $\dot{\theta}_i = \omega_i + \sum\limits_{j = 1}^{N}K_{ij}\sin(\theta_j-\theta_i)$   &  \makecell{Kuramoto model \\ $\m{x_i} := [\theta_i]$, $\theta_i$: oscillator phase}  \\ \midrule
				\textit{Epidemic outbreaks}\citep{LAJMANOVICH1976221}	& $\dot{p}_i = -\delta p_i + \sum\limits_{j = 1}^{N}a_{ij}\beta p_j(1-p_i)$ &  \makecell{$\m{x_i} := [p_i]$, $p_i\in[0,1]$ \\ $p_i$: infection probability}  \\ \midrule 
				\makecell{\textit{Mass-spring-damper} \\ \textit{systems}}	& \makecell{$m_i \ddot{x}_i = d_i\dot{x}_{i-1} -(d_i+d_{i+1})\dot{x}_{i}  +d_{i+1}\dot{x}_{i+1} + k_i{x}_{i-1} -(k_i+k_{i+1}){x}_{i}  +k_{i+1}{x}_{i+1}$ \\ $+ a_i{x}_{i-1}^3 -(a_i+a_{i+1}){x}_{i}^3  +a_{i+1}{x}_{i+1}^3\qquad\qquad\qquad\qquad\qquad\qquad\quad$ }  &  \makecell{$\m{x_i} := [x_i]$ \\ $x_i$: mass relative position 
				}  \\ \midrule 
				\bottomrule
		\end{tabular}}
	\end{center}
	\vspace{-0.6cm}
\end{table*}
\setlength{\floatsep}{10pt}

\vspace{-0.35cm}
\section{System Description and Preliminaries}\label{sec:model_problem}
{Consider a continuous-time networked NDS comprised of $N$ subsystems (also referred to as \textit{nodes}) modeled as} 
	\begin{align}
	\dot{\m x}(t) &= \mA \m x (t) + \m G\m f(\m x) + \mB \m u (t),\quad \m y(t) = \mC \m x (t).\label{eq:gen_dynamic_systems}
	\end{align}
{Numerous NDSs---see Table \ref{tab:nonlinear_models_cps} for some notable examples---can be represented in the form of \eqref{eq:gen_dynamic_systems}}. 
In this model, the global state $\m x\in \mathbfcal{X}\subset\mathbb{R}^{n_x}$ consists of $N$ numbers of nodal state $\m {x_i}\in \mathbfcal{X}_{\m i}\subset\mathbb{R}^{n_{x_i}}$. Likewise, each subsystem $\m i$ is also comprised of $n_{u_i}$inputs and $n_{y_i}$ measurements such that $\m {u_i}\in \mathbfcal{U}_{\m i}\subset\mathbb{R}^{n_{u_i}}$ and $\m {y_i}\in \mathbb{R}^{n_{y_i}}$. The notations $\m {u_i}$ and $\m {y_i}$ denote the nodal input and measurements vectors whereas $\mathbfcal{X}$ and  $\mathbfcal{U}$ represent the operating region and admissible inputs for NDS \eqref{eq:gen_dynamic_systems}. The global input and output vectors are $\m u\in \mathbfcal{U}\subset\mathbb{R}^{n_u}$ and $\m y\in \mathbb{R}^{n_y}$.
Matrices $\m A$, $\mB$, $\mC$, and $\mG$ are all assumed to have appropriate dimensions and in particular, $\mB$ and $\mC$ are representative of input-to-state and state-to-output mappings and supposed to posses the following structure: $\mB = \mathrm{Blkdiag}\left(\mB_{1},\hdots,\mB_N\right)$, $\mC = \mathrm{Blkdiag}\left(\mC_{1},\hdots,\mC_N\right)$. 
Matrix $\m A$ expresses the \textit{linear} dynamics of the system as well as interaction between subsystems while matrix  $\m G$ depicts the distribution of nonlinear mapping $\m f:\mathbb{R}^{n_x}\rightarrow \mathbb{R}^{n_g}$ that may capture any linear and {nonlinear} phenomena in the system.

The ultimate objective of this work is to search for both optimal and suboptimal SAs configurations or locations for NDSs (or equivalently, the number of nonzero columns of $\m B$ and nonzero rows of $\m C$)
while satisfying some user-defined constraints, 
which generally may include closed-loop system stability, convergence of estimation error dynamics, and actuator/sensor constraints. 
It is worth mentioning that this paper emphasizes solving SASPs for continuous-time NDSs with Lipschitz nonlinearities since \textit{(i)} Lipschitz is one of widely used function sets both in observer and controller designs---for example, see \cite{Phanomchoeng2010,nugroho2018journal}---and \textit{(ii)} the proposed methodology to solve SASPs can be directly extended for other classes of nonlinearities mentioned previously as well as for discrete-time NDSs with additional estimation/control objectives
To proceed, the subsequent assumption is considered in the paper.
\vspace{-0.1cm}
\begin{asmp}\label{def:lip}
	The mapping $\m f :\mathbb{R}^{n_x}\rightarrow \mbb{R}^{n_g}$ is locally Lipschitz continuous in $\mathbfcal{X}$ such that for any $\m x,\hat{\m x}\in \mathbfcal{X}$
	\begin{align}
	\norm{\m f(\m x)-\m f(\hat{\m x})}_2 \leq \gamma_l \norm{\m x - \hat{\m x}}_2,\label{eq:lip_def}
	\end{align}
	where $\gamma_{l} \geq 0$ is the corresponding Lipschitz constant.
\end{asmp}
\vspace{-0.1cm}
\noindent 
{Readers are referred to \citep{nugroho2020nonlinear} for scalable numerical methods to compute Lipschitz constants---including the ones corresponding to the other function sets.
In the next section, we formalize the notion of observability for Lipschitz NDSs which is crucial in the context of SSP.}

\vspace{-0.35cm}
\section{Observability for Lipschitz NDS}\label{sec:ssp}
In contrast to linear systems, quantifying observability for nonlinear systems is, without doubt, much more difficult and less straightforward. 
{To that end, in this section we focus on the concept of observability for NDS \eqref{eq:gen_dynamic_systems} in regard to solvability of the corresponding SDPs and Lipschitz constant.} To proceed, consider a NDS in the form of \eqref{eq:gen_dynamic_systems}.
A standard Luenberger-like observer for NDS \eqref{eq:gen_dynamic_systems} can then be constructed as follows
\vspace{-0.05cm}
\begin{subequations}\label{eq:obs_dynamic_systems_sasp}
	\begin{align}
	\hspace{-0.15cm}\dot{\hat{\m x}}(t) &= \mA \hat{\m x} (t) + \m G\m f(\hat{\m x}) + \mB \m u (t) + \m L (\m y(t)-\hat{\m y}(t))\\
	\hspace{-0.15cm}\hat{\m y}(t) &= \mC \hat{\m x} (t),
	\end{align}
\end{subequations}
where in \eqref{eq:obs_dynamic_systems_sasp}, $\m L\in\mbb{R}^{n_x\times n_y}$ is the observer gain matrix, $\hat{\m x}$ is the estimated states, and $\hat{\m y}$ is the estimated outputs. The objective here is to compute $\m L$ that makes the estimation error dynamics, where estimation error is constructed as $\m e(t):= \m x(t)-\hat{  \m x}(t)$, to converge asymptotically towards zero. The next definition quantifies observability for NDS \eqref{eq:gen_dynamic_systems}, posed as a SDP feasibility problem, for a given $\gamma_{l}$.
\vspace{-0.1cm}
\begin{mydef}\label{def:lip_observability}
	NDS \eqref{eq:gen_dynamic_systems} is said to be $\gamma_l$-observable if and only if there exist $\m P\in \mbb{S}^{n_x}_{++}$, $\m Y\in \mbb{R}^{n_x\times n_y}$, and $\epsilon\in\mbb{R}_{++}$ such that the following linear matrix inequality (LMI) is feasible
	\begin{align}
	\hspace{-0.6cm}\bmat{
		\m A ^{\top}\m P + \m P\m A - \m C ^{\top}\m Y ^{\top} -\m Y\m C  +\epsilon\gamma_l^2 \m I & * \\
		\m G ^{\top}\m P & -\epsilon \m I} & \prec 0. \label{eq:LMI_lip_obs}
	\end{align}
\end{mydef}  
\vspace{-0.05cm}   
LMI \eqref{eq:LMI_lip_obs} originates from \citep{Phanomchoeng2010} with special case $\m G = \m I$ and it can be proven that \eqref{eq:LMI_lip_obs} is indeed sufficient and necessary for the existence of stabilizing matrix $\m L$ for the estimation error dynamics, provided that $V(\m e) = \m e(t)^{\top}\mP \m e(t)$ is the Lyapunov function candidate and $\epsilon \in \mbb{R}_+$. Once the LMI is solved, $\m L$ can be recovered as $\mP^{-1}\mY$.	
The advantages of utilizing Definition \ref{def:lip_observability} to quantify observability are twofold. First, if there exists any solution for \eqref{eq:LMI_lip_obs} then one can immediately obtain stabilizing gain matrix $\m L$ and second, the observability prevails for various operating points in $\mathbfcal{X}$. These are in contrast to other types of observability for NDS such as empirical observability Gramian, as it only quantifies observability based on local behaviors of the systems around certain operating points and, upon determining the observability, one is still required to design a state estimator algorithm to estimate the actual states.

At this point, we are now compelled to consider the following question regarding the feasibility of SDP that corresponds to $\gamma_l$-observability for NDS \eqref{eq:gen_dynamic_systems}:

\vspace{0.1cm}

\noindent \textbf{Q1}: \textit{What role does constant $\gamma_l$ have on $\gamma_l$-observability?}

\vspace{0.1cm}

\noindent To answer \textbf{Q1}, we need to further study the LMI given in \eqref{eq:LMI_lip_obs}. {For the sake of simplicity, let us assume that $\epsilon > 0$ is a  constant. Then, realize that $\m Y\m C$ can be expressed as}
\vspace{-0.05cm}
\begin{align*}
\m Y\m C = \sum_{i=1}^{n_x}\sum_{j=1}^{n_y} Y_{ij}\check{\m C}_{ij}, 
\end{align*}
\vspace{-0.05cm}
where $\check{\m C}_{ij}\in \mbb{R}^{n_y\times n_x}$ is constructed as
\begin{align*}
\mathrm{row}\left(\check{\m C}_{ij}\right)_k = 
\begin{cases}
\left\{[C_{kl}]\right\}^{n_x}_{l = 1}\in \mbb{R}^{1\times n_x}, &\;\text{if}\;j = k\\
\m 0_{1\times n_x}, &\;\text{otherwise}.\\
\end{cases}
\end{align*}
Now let $\check{\m C}_{S,ij}:=  \check{\m C}_{ij} + \check{\m C}_{ij}^\top$ for each $i,j$.  
By using the above expression, the term $\m Y\m C + \m C ^{\top}\m Y ^{\top} $ can be expressed as
\vspace{-0.05cm}
\begin{align}
\m Y\m C + \m C ^{\top}\m Y ^{\top} 
&= \sum_{i=1}^{n_x}\sum_{j=1}^{n_y} Y_{ij}\check{\m C}_{ij} + \left(\sum_{i=1}^{n_x}\sum_{j=1}^{n_y} Y_{ij}\check{\m C}_{ij}\hspace{-0.05cm}\right)^{\hspace{-0.1cm}\top} \nonumber \\
&= \sum_{i=1}^{n_x}\sum_{j=1}^{n_y} Y_{ij}\check{\m C}_{S,ij} 
= \sum_{k=1}^{n_xn_y} y_{v,k}\bar{\m C}_k, \label{eq:CY_vec_LMI}
\end{align}
in which $\m y_v := [y_{v,1}\,\,y_{v,2}\,\,\cdots\,\,y_{v,n_xn_y}]^\top = \mathrm{vec}(\m Y)$ and $\bar{\m C}_k\in\mbb{S}^{n_x}$ for each $k\in\mbb{I}(n_xn_y)$ is associated with the corresponding $\check{\m C}_{S,ij}$.  
Using \eqref{eq:CY_vec_LMI}, LMI \eqref{eq:LMI_lip_obs} along with $\m P \succ 0$ can be written in the following compact form
\begin{subequations}\label{eq:obs_LMI_ref}
	\vspace{-0.05cm}
	\begin{align}
	\mathcal{A}(\m P, \m y_v)  + \mathcal{A}_0 \succ 0,\label{eq:obs_LMI_ref_1}
	\end{align}  
	in which $\mathcal{A}: \mbb{S}^{n_x}\times \mbb{R}^{n_xn_y}\rightarrow \mbb{S}^{2n_x}\times\mbb{S}^{n_x}$ is constructed as $\mathcal{A}(\m P, \m y_v) := \mathcal{A}_{P}(\m P) + \mathcal{A}_{Y}({\m y_v})$ and
	\begin{align}
	\mathcal{A}_{P}(\m P) &:= \mathrm{Blkdiag}\left(-\bmat{\m A ^{\top}\m P + \m P\m A & * \\ \m G ^{\top}\m P & \m O}, \m P\right)\label{eq:obs_LMI_ref_2}\\
	\mathcal{A}_{Y}({\m y_v}) &:= \mathrm{Blkdiag}\left(\bmat{\sum_{i=1}^{n_xn_y} y_{v,i}\bar{\m C}_i & *\\ \m O&\m O},\m O\right)\label{eq:obs_LMI_ref_3}\\
	\mathcal{A}_0 &:= \mathrm{Blkdiag}\left(\bmat{-\epsilon\gamma_l^2 \m I & *\\ \m O&\epsilon \m I},\m O\right).\label{eq:obs_LMI_ref_4}
	\end{align}
\end{subequations}
Using the expression provided in \eqref{eq:obs_LMI_ref}, we now present the first theoretical result of the paper.
\vspace{-0.15cm} 
\begin{theorem}[Larger Lipschitz Constant Reduces Observability]\label{thm:lip_obs_lmi}
	There exist $\m P\in \mbb{S}^{n_x}$ and $\m y_v\in \mbb{R}^{n_xn_y}$ such that $\mathcal{A}(\m P, \m y_v)  + \mathcal{A}_0\succ 0$ if and only if there is no $\m Z\in \mbb{S}^{3n_x}$ in the form of
	\begin{subequations}\label{eq:lip_obs_lmi_thm}
		\begin{align}
		\m Z = \mathrm{Blkdiag}\left(\bmat{\m Z_1& * \\ \m Z_2^\top & \m Z_3}, \m Z_4\right),\label{eq:lip_obs_lmi_thm_1}
		\end{align}
		such that $\m Z \succeq 0$ and $\m Z \neq 0$ satisfying
		\begin{align}
		&\m Z_4 = \m Z_1\m A^\top + \m A\m Z_1 + \m G\m Z_2^\top + \m Z_2\m G^\top \label{eq:lip_obs_lmi_thm_2}\\
		&\sum_{j=1}^{n_x}\sum_{k=1}^{n_x}\bar{C}_{i,jk} Z_{1,jk}  = 0,\;\;\forall i\in\mbb{I}(n_xn_y) \label{eq:lip_obs_lmi_thm_3}\\
		&\mathrm{tr}(\m Z_3) \leq \gamma_l^2 \,\mathrm{tr}(\m Z_1).	\label{eq:lip_obs_lmi_thm_4}	
		\end{align}
	\end{subequations}
\end{theorem}
\vspace{-0.1cm}
\noindent The proof of the above theorem is presented in \ref{appdx:B}. Problem \eqref{eq:lip_obs_lmi_thm} is referred to throughout the section as the \textit{alternative problem} for \eqref{eq:obs_LMI_ref_1}. To answer \textbf{Q1}, we need to look into inequality \eqref{eq:lip_obs_lmi_thm_4}.
Realize that the term $\mathrm{tr}(\m Z_3)$ is lower bounded by zero since $\m Z\succeq 0$. Consequently, we get
\vspace{-0.05cm}
\begin{align}
0 \leq \mathrm{tr}(\m Z_3) &\leq \gamma_l^2 \,\mathrm{tr}(\m Z_1).	\label{eq:lip_obs_lmi_infeasibility}	
\end{align}
It is seen from \eqref{eq:lip_obs_lmi_infeasibility} that the feasible set of \eqref{eq:lip_obs_lmi_thm} expands as the value of $\gamma_l$ increases. 
In a particular case when $\gamma_l = 0$, the term $\mathrm{tr}(\m Z_3)$ is enforced to be equal to zero. Since eigenvalues of $\m Z_3$ cannot be negative, then they must be equal to zero. On the other hand, if $\gamma_l$ is considerably large, then there is not much restriction on the upper bound of $\mathrm{tr}(\m Z_3)$. Hence, it is important to get smaller Lipschitz constant $\gamma_l$ since this will make the problem described in \eqref{eq:lip_obs_lmi_thm} to have narrower feasible space. 
{This finding is in accordance with empirical observations from the literature: SDP \eqref{eq:LMI_lip_obs} has no solution if $\gamma_l$ is selected to be sufficiently large---such large Lipschitz constant is labeled as \textit{conservative}. This provides an answer for \textbf{Q1}. 
The next section introduces the SSP for Lipschitz NDSs and presents some properties useful for the development of a more efficient BnB algorithm.}

\vspace{-0.35cm}
\section{Sensor Selection and NDS Observability}\label{sec:obs_ssp_sensor}
{In this section, we wish to answer the question below:}

\vspace{0.05cm}

\noindent \textbf{Q2}: \textit{What is the impact of the number of placed sensors on $\gamma_l$-observability?}

\vspace{0.05cm}

The above question may seem trivial since it is expected that dynamic systems have \textit{higher degree of observability} as more sensors are placed or utilized. The objective of this section is to demonstrate the above conjecture from $\gamma_l$-observability stand point. To begin with, we formally construct the SSP for NDS \eqref{eq:gen_dynamic_systems} as follows. Let $\gamma_i\in\{0,1\}$ be a binary variable that ascertains the activation or deactivation of sensor on each subsystem $i$. That is, $\gamma_i = 1$ if the sensor measuring subsystem $i$ is activated and $\gamma_i = 0$ otherwise. These variables can be combined into $\m \gamma$, giving $\m \gamma := \left[\gamma_1\,\,\gamma_2\,\,\cdots\,\,\gamma_N\right]^{\top}$.
The following conventions are adopted in this section.
\vspace{-0.1cm} 
{\begin{mydef}~\label{def:setS}
	Let $\mathcal{S}_\gamma$ be an $N$-tuple representing the selection of sensors, i.e., $\mathcal{S}_\gamma := (\gamma_1,\gamma_2,\ldots,\gamma_N)$. 
	The set of active sensor is constructed as $\mathcal{S}_{\gamma}^{(a)} := \left\{\gamma \in \mathcal{S}_\gamma\,|\,\gamma =1\right\}$. 
	Let $\bar{\mathcal{S}}_\gamma := (\gamma_1\m 1_{n_{y_1}},\gamma_2\m 1_{n_{y_2}},\ldots,\gamma_N\m 1_{n_{y_N}})$ where $\mathrm{card}(\bar{\mathcal{S}}_\gamma) = n_y$. The selection of sensors in matrix form can be written as $\m \Gamma := \mathrm{Blkdiag}\big(\gamma_1\m I_{n_{y_1}},\gamma_2\m I_{n_{y_2}},\hdots,\gamma_N\m I_{n_{y_N}}\big)$.
\end{mydef}}
\vspace{-0.1cm} 
By using the above definition, NDS \eqref{eq:gen_dynamic_systems} together with sensor selection can be conveniently expressed as
\vspace{-0.0cm}
\begin{subequations}\label{eq:gen_dynamic_systems_sen_sel}
	\begin{align}
	\dot{\m x}(t) &= \mA \m x (t) + \m G\m f(\m x) + \mB\m u(t)\\
	\m y(t) &= \m \Gamma\mC \m x (t).
	\end{align}
\end{subequations}
Additionally, it is also beneficial to consider a set $\mathcal{G}_\gamma\subseteq \{0,1\}^N$ representing logistic constraints and availability of sensors such that $\m \gamma \in \mathcal{G}_\gamma$ might be imposed.
{This in turn allows a simplified high level formulation of a general SSP} 
\vspace{-0.05cm}
\begin{align*}
\mathbf{(P1)}\;\;\;	\minimize \;\;\;&  \m c^{\top} \boldsymbol \gamma + \mathrm{EstimationObjective} \\
\subjectto  \;\;\;& \eqref{eq:gen_dynamic_systems_sen_sel},\; \m \gamma\in \mathcal{G}_\gamma,\; \mathrm{EstimationConstraints}. 
\end{align*}
\vspace{-0.05cm}
The objectives of \textbf{P1} are threefold: \textit{(i)} performing state estimation for NDS \eqref{eq:gen_dynamic_systems_sen_sel} while \textit{(ii)} utilizing smallest number of sensors as possible (or satisfying a given constraint over the collections of library of sensors) and \textit{(iii)} optimizing a specific estimation metric. Vector $\m c\in\mbb{R}^N_{+}$ in the objective function of \textbf{P1} assigns weights for each sensor $\gamma_i$. Given \textbf{P1}, SSP for Lipschitz NDSs assuming no estimation objective (this is considered in Section~\ref{sec:misc}), can be constructed by incorporating \eqref{eq:gen_dynamic_systems_sen_sel} into LMI \eqref{eq:LMI_lip_obs}. The resulting problem is given below.
\vspace{-0.05cm} 
\begin{subequations}\label{eq:sen_sel_obs}
	\begin{align}
&\hspace*{-0.2cm}\mathbf{(P2)}\;	\minimize_{ \m P, \m Y, \epsilon, \m \gamma}\;\;  \m c^{\top} \boldsymbol \gamma \\
	&\hspace*{-0.2cm}\subjectto  \;\;	  \begin{bmatrix}
	\m A ^{\top}\m P + \m P\m A   +\epsilon\gamma_l^2\m I  & \\ 
	- \m Y\m \Gamma \m C- \m C ^{\top}\m \Gamma\m Y ^{\top}& *\\
	\m G ^{\top}\m P & -\epsilon \m I \end{bmatrix} \preceq 0 \label{eq:sen_sel_obs_1} \\ 
	&\quad \quad\;\;\m P\succ 0, \;\epsilon > 0,\; \m \gamma\in \mathcal{G}_\gamma,\;
	\m \gamma \in \{0,1\}^N.\label{eq:sen_sel_obs_2}\vspace*{-0.05cm}
	\end{align}
\end{subequations}
\vspace{-0.05cm} 
In \textbf{P2} the objective is to minimize the number of the activated sensors while \textit{(a)} finding a stabilizing observer gain matrix $\mL$ and \textit{(b)} satisfying sensor's logistic constraints. As $\m \Gamma$ is a binary matrix variable, \textbf{P2} is a nonconvex optimization problem with \textit{mixed-integer bilinear matrix inequalities} (MIBMIs) because of the $\m Y \m \Gamma$ term. Our approach to address this problem is given in the next section. {Following \eqref{eq:CY_vec_LMI}, the term $ \m Y\m \Gamma \m C+\m C ^{\top}\m \Gamma\m Y ^{\top}$ is equal to
\vspace{-0.05cm} 
\begin{align}
\sum_{j=1}^{n_y} \gamma_j\left(\sum_{i=1}^{n_x} Y_{ij}\check{\m C}_{S,ij}\right) = \sum_{k=1}^{n_xn_y} \bar{\gamma}_{k} y_{v,k}\bar{\m C}_k, \label{eq:CGammaY_vec_LMI}
\end{align} }
\vspace{-0.05cm} 
\noindent where each $\bar{\gamma}_{k}$ corresponds to ${\gamma}_{j}\in \bar{\mathcal{S}}_\gamma$. As such we may write $\bar{\gamma}_{k}\triangleleft{\gamma}_{j}$. For the sake of analysis, let us assume that $\m\gamma$ is fixed. From \eqref{eq:CGammaY_vec_LMI}, define $\mathcal{A}_{Y,\Gamma}({\m y_v})$ as follows
\begin{subequations}
	\begin{align}
	\hspace{-0.3cm}\mathcal{A}_{Y,\Gamma}({\m y_v}) &:= \mathrm{Blkdiag}\left(\bmat{\sum_{k=1}^{n_xn_y}\bar{\gamma}_{k} y_{v,k}\bar{\m C}_k & *\\ \m O&\m O},\m O\right), \label{eq:obs_LMI_ref_gamma}
	\end{align} 
	such that, by having $\mathcal{A}_{\Gamma}(\m P, \m y_v) := \mathcal{A}_{P}(\m P) + \mathcal{A}_{Y,\Gamma}({\m y_v})$, matrix inequality \eqref{eq:obs_LMI_ref_1} now becomes
	\begin{align}
	\hspace{-0.4cm}\mathcal{A}_{\Gamma}(\m P, \m y_v) + \mathcal{A}_0 = \mathcal{A}_{P}(\m P) + \mathcal{A}_{Y,\Gamma}({\m y_v})+ \mathcal{A}_0 \succ 0.\label{eq:obs_LMI_ref_gamma_problem}
	\end{align}
\end{subequations}
The next result is established due to Theorem \ref{thm:lip_obs_lmi}. 
\vspace{-0.05cm} 
\begin{theorem}\label{thm:lip_obs_lmi_sensor}
	Consider a certain sensor configuration $\mathcal{S}_\gamma$. There exist $\m P\in \mbb{S}^{n_x}$ and $\m y_v\in \mbb{R}^{n_xn_y}$ satisfying $\mathcal{A}_{\Gamma}(\m P, \m y_v)  + \mathcal{A}_0\succ 0$ if and only if there is no $\m Z\in \mbb{S}^{3n_x}$ provided in the form of \eqref{eq:lip_obs_lmi_thm_1}
	such that $\m Z \succeq 0$ and $\m Z \neq 0$ satisfying \eqref{eq:lip_obs_lmi_thm_2}, \eqref{eq:lip_obs_lmi_thm_4}, and $\m Z_1\in \mathcal{Z}\big(\mathcal{S}_{\gamma}^{(a)}\big)$ where 
	\begin{align}
\hspace{-0.2cm}	\mathcal{Z}\big(\mathcal{S}_{\gamma}^{(a)}\big) := &\Bigg\{\vphantom{\Bigg\vert}\m Z_1\in \mbb{S}^{n_x}\,\Bigg\vert\,
	\sum_{j=1}^{n_x}\sum_{k=1}^{n_x}\bar{\gamma}_{i}\bar{C}_{i,jk} Z_{1,jk}  = 0, \nonumber \\
\hspace{-0.2cm}		&\quad \;\hspace*{-0.2cm}\;\forall i\in\mbb{I}(n_xn_y)\;\wedge\;\bar{\gamma}_{i}\triangleleft{\gamma}_{m},\;\gamma_m\in\mathcal{S}_{\gamma}^{(a)}\vphantom{\Bigg\vert}\Bigg\}.\label{eq:lip_obs_lmi_sensor}
	\end{align}
\end{theorem}
\vspace{-0.1cm}
{It follows from Theorem \ref{thm:lip_obs_lmi_sensor} (see \ref{appdx:C} for the proof) that including more sensors will reduce the feasible set of the alternative problem and thus, in general, utilizing more sensors reduces the feasible set of the alternative problem.} By reducing this set, one can expect that the constraint will be inconsistent thus making the alternative problem infeasible and consequently, giving a feasible solution for \eqref{eq:obs_LMI_ref_gamma_problem}. This provides an answer for \textbf{Q2}. {From Theorem \ref{thm:lip_obs_lmi_sensor}, the following additional result is obtained.}
\vspace{-0.1cm}
\begin{mycor}\label{cor:lip_obs_lmi_sensor}
	Suppose $\mathcal{S}_{\gamma,1}$ and $\mathcal{S}_{\gamma,2}$ are two distinct sensor's configurations so that $\mathcal{S}_{\gamma,1}^{(a)}\hspace{-0.05cm}\subset \mathcal{S}_{\gamma,2}^{(a)}$. If \eqref{eq:obs_LMI_ref_gamma_problem} has no solution for $\mathcal{S}_{\gamma,2}$, then it also has no solution for $\mathcal{S}_{\gamma,1}$. 
\end{mycor}
\vspace{-0.1cm}
\noindent The above corollary essentially states that if the system is not $\gamma_l$-observable for a particular sensor configuration  $\mathcal{S}_{\gamma,2}$, then the system is also not $\gamma_l$-observable for any sensor configuration, represented by $\mathcal{S}_{\gamma,1},$ having fewer number of sensors if $\mathcal{S}_{\gamma,1}^{(a)}\hspace{-0.05cm}\subset \mathcal{S}_{\gamma,2}^{(a)}$. 
The proof of Corollary \ref{cor:lip_obs_lmi_sensor} follows directly from the fact that $\mathcal{S}_{\gamma,1}^{(a)}\subset \mathcal{S}_{\gamma,2}^{(a)}$ implies $\mathcal{Z}\big(\mathcal{S}_{\gamma,1}^{(a)}\big) \subset\mathcal{Z}\big(\mathcal{S}_{\gamma,2}^{(a)}\big)$. That is, since the alternative problem with set of constraints $\mathcal{Z}\big(\mathcal{S}_{\gamma,2}^{(a)}\big)$ has nonempty feasible set, then this set is never empty for any $\mathcal{S}_{\gamma,1}$ since $\mathcal{Z}\big(\mathcal{S}_{\gamma,1}^{(a)}\big)$ is less constrained. 
By contrapositive, then it is also true that the system is $\gamma_l$-observable for any sensor configuration  $\mathcal{S}_{\gamma,2}$ provided that it is also $\gamma_l$-observable for a certain sensor configuration $\mathcal{S}_{\gamma,1}$ such that $\mathcal{S}_{\gamma,1}\hspace{-0.05cm}\subset \mathcal{S}_{\gamma,2}$. 

Now, a natural question to ask is whether there is a relation between the value of Lipschitz constant $\gamma_l$ and the number of activated sensors to achieve $\gamma_l$-observability for SSP. Consider a SSP where the objective is to achieve $\gamma_l$-observability with minimum number of activated sensors. From previous results, we know that minimizing $\gamma_l$ while incorporating more sensors reduces the feasible set of the alternative problem. However, as we want to incorporate the least number of sensors as possible, then we can only rely on using the smallest Lipschitz constant $\gamma_l$. The rationale here is, if the alternative problem can be made infeasible from \eqref{eq:lip_obs_lmi_infeasibility} by reducing the Lipschitz constant, then we can utilize less number of activated sensors to achieve $\gamma_l$-observability. 
In the next section, we discuss our approach to solve SSP formulated in \textbf{P2}.    

\vspace{-0.25cm}
\section{Solving the SSP Through Customized BnB}\label{sec:sol_appr}


\vspace{-0.25cm}
\subsection{From MIBMI to Convex MISDP}\label{ssec:mibmi_to_misdp}

Our approach involves transforming the SSP to a convex MISDP. 
Specifically, our prior work \citep{Nugroho2019CDC} reformulates the SSP \textbf{P2} into a covex MISDP using McCormick's relaxation \citep{mccormick1976computability}. This is not the only method to convert MIBMI into MISDP: the big-M method, which is popular in disjunctive programming, can also be employed---see \citep{Nugroho2019124}.
The McCormick's reformulation is performed by defining a new matrix variable $\m M := \m Y \m \Gamma$ where $\m M\in\mbb{R}^{n_x\times n_y}$ given that $\m Y$ is bounded such that $\barbelow{\m Y} \leq \m Y \leq \bar{\m Y}$, while the big-M assumes that $-L \m 1 \leq \m Y \leq L \m 1$ for a sufficiently large constant $L > 0$. For convenience, we consider $\m Y \in B(\m Y)$ where $B(\m Y)$ is defined as
\begin{align*}
B(\m Y) := \begin{cases}
\barbelow{\m Y} \leq \m Y \leq \bar{\m Y}, &\;\text{for McCormick},\\
-L \m 1 \leq \m Y \leq L \m 1, &\;\text{for Big-M}.\\
\end{cases}
\end{align*} 
It is apparent here that the bounds on $\m Y$ are determined by a single constant $L$ for big-M and matrices $\barbelow{\m Y},\bar{\m Y}$ for McCormick, making big-M to be a special case of McCormick's relaxation.  
The resulting problem can now be written as
\vspace{-0.0cm}
\begin{subequations}\label{eq:sen_sel_obs_const}
	\begin{align}
	&\hspace*{-0.2cm}(\mathbf{P3})\;	\minimize_{ \m P, \m Y, \m M, \epsilon, \m \gamma}\;\;  \m c^{\top} \boldsymbol \gamma \\
	&\hspace*{-0.2cm}\subjectto  \;\;	  \begin{bmatrix}
	\m A ^{\top}\m P + \m P\m A   +\epsilon\gamma_l^2\m I  & \\ 
	- \m M \m C- \m C ^{\top}\m M^{\top}& *\\
	\m G ^{\top}\m P & -\epsilon \m I \end{bmatrix} \preceq 0 \label{eq:sen_sel_obs_const_1} \\ 
	&\quad \quad \quad \quad\;\;\m M = \m Y \m \Gamma, \;\m Y \in B(\m Y),\;\eqref{eq:sen_sel_obs_2}.\label{eq:sen_sel_obs_const_2}\vspace*{-0.05cm}
	\end{align}
\end{subequations}
Without loss of generality, it is assumed throughout the section that the number of variables in $\m \gamma$ are equal to the number of rows in $\m C$.
By applying big-M or McCormick's relaxation, the nonconvex MISDP \textbf{P3} can be converted to an equivalent convex MISDP given in the next theorem.
\vspace{-0.1cm}
\begin{theorem}\label{thm:misdp}
	Problem \textbf{P3} is equivalent to 
	\begin{subequations}\label{eq:sen_sel_misdp_mccormick}
		\vspace*{-0.00cm}
		\begin{align}
		\hspace*{-0.1cm}\mathbf{(P4)}\;	\minimize_{\m v}\;\; & \m c^{\top}_v \m v \label{eq:sen_sel_misdp_mccormick1}\\
		\subjectto  \;\;	 & \mathcal{L}_v(\m v) + \mathcal{C}_v\succeq 0 , \label{eq:sen_sel_misdp_mccormick2}\\
		\;\;& v_i\in \{0,1\},\;\;\forall i\in\mbb{I}_v(\m \gamma), \label{eq:sen_sel_misdp_mccormick3}
		\end{align}
		where $\mathcal{L}_v(\cdot)$ denotes LMI terms, $\mathcal{C}_v$ denotes the constant terms, and vector $\m v\in\mbb{R}^{n_v}$ populates all decision variables in the following fashion 
		\begin{align}
		\m v := \bmat{\mathrm{vec}(\m P)^{\top}\;\;\epsilon\;\;\mathrm{vec}(\m M)^{\top}\;\;\mathrm{vec}(\m Y)^{\top}\;\;\m \gamma^{\top}}^{\top}, \label{eq:vector_v}
		\end{align}
	\end{subequations}     	
	with $n_v = {\bar{n}_x+1+2n_xn_y+n_y}, \bar{n}_x := \tfrac{1}{2}n_x(n_x+1)$.
\end{theorem}

{The proof of Theorem \ref{thm:misdp} is detailed in \ref{appdx:Z}}.
The notation  $\mbb{I}_v(\m \gamma)$ in \eqref{eq:sen_sel_misdp_mccormick3} is useful for indicating the index $i$ of $\m v$ such that $v_i = \gamma_j$ for each corresponding $j\in\mbb{I}(\m \gamma)$; similar notations are also used to label any variable in $\m v$ that corresponds to $\m P$, $\m Y$, $\m M$, and $\epsilon$.
Specifically, the inequality $\m \Xi \,\m \nu \leq \m \varphi$ which is embedded in LMI \eqref{eq:sen_sel_misdp_mccormick2} essentially encompasses the McCormick's envelope on the bilinear equality $M_{ij} = Y_{ij}\gamma_j$ expressed as
\begin{align}
\bmat{1&-1&-\barbelow{Y}_{ij}\\ -1&1&\bar{Y}_{ij}\\1&0&-\bar{Y}_{ij}\\-1&0&\bar{Y}_{ij}}\bmat{M_{ij}\\Y_{ij}\\\gamma_j} \leq \bmat{-\barbelow{Y}_{ij} \\ \bar{Y}_{ij}\\0\\0}, \label{eq:mccormick_envelope}
\end{align} 
which is equivalent to the previous bilinear constraint (similarly for big-M).
Now that we have a convex MISDP, problem \textbf{P4} can be solved using one of the aforementioned techniques. In the next section, we develop a customized BnB algorithm that can be used to compute both optimal and suboptimal solutions for \textbf{P4}.

\vspace{-0.25cm}
\subsection{Structure-Exploiting BnB Algorithm }\label{ssec:bnb_misdp}
A natural way to solve such MISDP is through BnB algorithm. In a general purpose BnB algorithm, an SDP is solved in every node, in which the SDP is obtained from relaxing the integer variables to their corresponding convex relaxations. Thus, the complexity of BnB algorithm stems from the difficulties in solving the SDPs and the number of SDPs that need to be solved. 
In a practical situation, finding a relatively good solution in a much shorter time is more desirable than finding an optimal one. 
{Despite that some general purpose BnB algorithms for MISDPs are available,
these conventional solvers cannot be configured to make them amenable for exploiting the structure of the problem at hand in order to reduce the complexity of solving SSP.}       

\setlength{\textfloatsep}{5pt}
{\begin{algorithm}[t]
		\caption{\text{SE-BnB Algorithm}}\label{alg:BnB-SSP}
		\DontPrintSemicolon 
		\textbf{input:} $\mathcal{F}$, $\epsilon_f$, $\mathrm{maxBranch}$\;
		\textbf{initialize:} $\mathcal{C} = \emptyset$, $\mathcal{W} = \emptyset$, $\mathrm{LB},\mathrm{UB} = \infty$, $k = 0$, $\m v^* = \m 0$\;
		\textbf{process root node:} $\mathcal{T}\leftarrow\mathrm{RelaxSol}(\mathcal{F})$\;
		\textbf{update:}  $\mathrm{LB}\leftarrow \m c_v^{\top}\mathcal{T}(2)$, 
		$\mathcal{C}\leftarrow\mathcal{C}\cup\{\mathcal{T}\}$\;
		\While{$ \mathrm{UB}-\mathrm{LB}> \epsilon_f$ {\bf and}   $k < \mathrm{maxBranch}$\label{alg:1_loop}}{
			$\tilde{\mathcal{T}} = \argmin_{\mathcal{T}_k\in \mathcal{C}}  \m c_v^{\top}\mathcal{T}_k(2)$, 
			$\mathcal{C}\leftarrow\mathcal{C}\setminus\{\tilde{\mathcal{T}}\}$, $\mathrm{LB}\leftarrow \m c_v^{\top}\tilde{\mathcal{T}}(2)$\;
			\If{$\tilde{\mathcal{T}}(2)_i\in \{0,1\},\;\forall i\in\mbb{I}_v(\m \gamma)$}{
				$\tilde{\mathcal{T}}(3)\leftarrow\tilde{\mathcal{T}}(2)$\;
			}\Else{
				$\{\tilde{\mathcal{T}}(3),\mathcal{W}\}\leftarrow\mathrm{FeasSol}(\tilde{\mathcal{T}},\mathcal{W})$\;
				\textbf{find:} $i\in\mbb{I}_v(\m\gamma)$ such that $\tilde{\mathcal{T}}(2)_i\notin \{0,1\}$ and $\tilde{\mathcal{T}}(2)_i = \min_{j\in\mbb{I}_v(\m\gamma)} \abs{\tilde{\mathcal{T}}(2)_j-\tfrac{1}{2}}$\;
				$\mathcal{F}_L:= \{\m v\in \tilde{\mathcal{T}}(1)\,\vert \,v_i\leq 0\}$ \;
				\If{$\forall\,\m \gamma\in \{0,1\}^{n_y}, \m \gamma \in \mathbfcal{I}(\mathcal{F}_L)$, then $\m \gamma \notin \mathcal{W}$\label{step:1a}}{	$\mathcal{T}_L\leftarrow\mathrm{RelaxSol}(\mathcal{F}_L)$, $\mathcal{C}\leftarrow\mathcal{C}\cup\{\mathcal{T}_L\}$}
				$\mathcal{F}_R:= \{\m v\in \tilde{\mathcal{T}}(1)\,\vert\, v_i\geq 1\}$\;
				\If{$\forall\,\m \gamma\in \{0,1\}^{n_y}, \m \gamma \in \mathbfcal{I}(\mathcal{F}_R)$, then $\m \gamma \notin \mathcal{W}$\label{step:1b}}{$\mathcal{T}_R\leftarrow\mathrm{RelaxSol}(\mathcal{F}_R)$, $\mathcal{C}\leftarrow\mathcal{C}\cup\{\mathcal{T}_R\}$}
				$k\leftarrow k+1$\;
			}
			\If{$\m c_v^{\top}\tilde{\mathcal{T}}(3) < \mathrm{UB}$}{$\mathrm{UB}\leftarrow\m c_v^{\top}\tilde{\mathcal{T}}(3)$, $\m v^* \leftarrow \tilde{\mathcal{T}}(3)$}
			\ForEach{$\mathcal{T}_j\in C$}{
				\If{$\m c_v^{\top}\mathcal{T}_j(2) > \mathrm{UB}$}{%
					$\mathcal{C}\leftarrow \mathcal{C}\setminus \{\mathcal{T}_j\}$\;
				}
			}
		}
		\textbf{output:} $\mathrm{LB}$, $\mathrm{UB}$, $\m v^*$\;
	\end{algorithm}
}

To that end, we present a customized BnB algorithm that can be utilized to compute optimal and suboptimal solutions for \textbf{P4} in a more efficient manner, owing it to \textit{(a)} its ability to exploit problem structure and \textit{(b)} heuristics to quickly find good suboptimal solutions.  
To proceed, define
\begingroup
\allowdisplaybreaks
\begin{align*}
\mathcal{F} &:= \{\m v\,\vert\,\mathcal{L}_v(\m v) + \mathcal{C}_v\succeq 0,\;v_i\in \{0,1\},\;\forall i\in\mbb{I}_v(\m \gamma)\} \\
\mathcal{R} &:= \{\m v\,\vert\,\mathcal{L}_v(\m v) + \mathcal{C}_v\succeq 0,\;v_i\in [0,1],\;\forall i\in\mbb{I}_v(\m \gamma)\}, 
\end{align*} 
\endgroup
i.e., $\mathcal{F}$ denotes the feasible set for \textbf{P4} whereas $\mathcal{R}$ is its convex relaxation.  
Note that we have $\mathcal{F} \subseteq\mathcal{R}$ because in $\mathcal{R}$, the integer constraints are relaxed. 
The proposed BnB algorithm solves two convex problems on each BnB node in order for it to be able to find a good suboptimal solution immediately: the convex relaxation of \textbf{P4} to get a \textit{local} lower bound, and \textbf{P4} with fixed SAs combination to get a \textit{global} upper bound. 
Within this section, it is supposed that each node $i$ is representable by a $3$-tuple $\mathcal{T}_i := \left(\mathcal{F}_i,\barbelow{\m v},\bar{\m v}\right)$, where
\begin{align*}
\barbelow{\m v} = \argmin_{\m v\in\mathcal{R}_i}\m c_v^{\top}{\m v},\;\quad\;\bar{\m v}\in \{\m v\;\vert\; \m v\in\mathcal{F}_i\},
\end{align*}
such that $\mathcal{F}_i = \mathcal{T}_i(1)$, $\barbelow{\m v} =  \mathcal{T}_i(2)$, and $\bar{\m v} =  \mathcal{T}_i(3)$.
We also use an abstract set $\mathcal{I}(\cdot)\subseteq \mbb{R}^{n_y}$ where $\mathcal{I}(\mathcal{T}_i) \subset \mathcal{T}_i(1)$ to represent the restrictions (or \textit{integer cut}) on integer variable $v_i$, where $i\in \mbb{I}_v(\m \gamma)$, which is added in the branching process of BnB algorithm.  
To stop the algorithm when a suboptimal solution is sought, two stopping criteria characterize by $\epsilon_f\in\mbb{R}_+$ and $\mathrm{maxBranch}\in\mbb{N}_+$ are employed. Let $\mathrm{LB}$ and $\mathrm{UB}$ be the known best lower and upper bounds at the end of the algorithm. When the optimality gap $\mathrm{UB}-\mathrm{LB}$ is zero, the solution is optimal. However, if $\mathrm{UB}-\mathrm{LB} \leq \epsilon_f$ the solution is regarded to be $\epsilon_f$-optimal. Alternatively, the solution is only suboptimal. The notation $\mathrm{maxBranch}$, sometimes abbreviated as $\mathrm{mBr}$, determines the number of maximum allowable branching. These BnB routines,  referred to as \textit{structure-exploiting BnB} (SE-BnB), are detailed in Algorithm \ref{alg:BnB-SSP}. The algorithmic function $\mathrm{RelaxSol}(\cdot)$ solves the relaxed problem $\barbelow{f} = \min_{\m v\in\mathcal{R}}c_v^{\top}{\m v}$ for a given $\mathcal{F}$ where $\barbelow{\m v}$ is the minimizer and wrap the results into $\mathcal{T}$. 
In addition to this, Algorithm \ref{alg:BnB-SSP} uses a heuristic procedure for efficiently finding a feasible solution for \textbf{P4} on each node, if such solution exists. This heuristic generates a random combination of sensors $\m \gamma_c$. To do so, we define the set $\mathcal{W}$ as 
\vspace{-0.05cm}
\begin{align*}
\mathcal{W} := \left\{\m\gamma\in\{0,1\}^{n_y}\,\Big\vert\,\bmat{\m w^{\top}\,\,\m\gamma^{\top}}^{\top}\notin \mathcal{F},\;\forall \m w\right\},
\end{align*} 
i.e., the set of combinations of sensors that are infeasible for \textbf{P4}. 
By exploiting the property proposed in Corollary \ref{cor:lip_obs_lmi_sensor}, we are able to effectively generate a candidate of sensor combination $\m \gamma_c$ that is not positively infeasible for \textbf{P4}.
This heuristic, embedded in the function $\mathrm{FeasSol}(\cdot)$, is presented in Algorithm \ref{algo_feassol}. 
This idea is implemented in Steps \ref{step:1a} and \ref{step:1b} of Algorithm \ref{alg:BnB-SSP}.
Now, some modifications for making the SE-BnB algorithm to be more efficient are discussed. 

\setlength{\floatsep}{5pt}
{
	\begin{algorithm}[t]
		\caption{$\mathrm{FeasSol}$}\label{algo_feassol}
		\DontPrintSemicolon
		\textbf{input:} $\tilde{\mathcal{T}}$, $\mathcal{W}$\;
		\textbf{candidate generation:} generate a random sensor combination $\m \gamma_c \in \{0,1\}^{n_y}$ such that $\m \gamma_c\notin \mathcal{W}$ and
		$\m \gamma_c \in \mathcal{I}(\tilde{\mathcal{T}}(1))$\;
		\textbf{solve:} $\bar{f} = \min_{\m v\in\tilde{\mathcal{T}}(1),\,v_i = \gamma_{c,i}}c_v^{\top}{\m v}$ (problem \textbf{P6}) with minimizer $\bar{\m v}$\;
		\If{$\bar{f} = \infty$}{
			$\mathcal{W}\leftarrow \mathcal{W}\cup \{\m\gamma_c\}$
		}
		\textbf{output:} $\bar{\m v}$, $\mathcal{W}$
\end{algorithm}}

\vspace{-0.2cm}
\begin{itemize}[leftmargin=*]
		\setlength\itemsep{-0.3em}
	\item In $\mathrm{RelaxSol}(\cdot)$, the relaxed \textbf{P4} is solved. Realize that an integer cut is appended in every node (apart from the root node) that corresponds to whether a particular relaxed integer variable, say $v_i$ for $i\in\mbb{I}_v(\m\gamma)$, is constrained to be $v_i \leq 0$ or  $v_i \geq 1$. As $v_i\in[0,1]$, then the only feasible solution is  $v_i = 0$ or $v_i = 1$. Either case, the resulting McCormick's envelope (or big-M) yields the \textit{best} relaxation which makes \eqref{eq:mccormick_envelope} redundant. At this instance, constraint \eqref{eq:mccormick_envelope} can be discarded and replaced with $M_{ij} = 0$ or $M_{ij} = Y_{ij}$, depending on the value of $v_i$. 
	\item Recall that the SE-BnB algorithm attempts to find a feasible solution on each node by fixing $\m \gamma$ and solve \textbf{P4}---performed in $\mathrm{FeasSol}(\cdot)$. Since \textbf{P4} is equivalent to \textbf{P3} and \textbf{P3} becomes a SDP when $\m \gamma$ is fixed, then one can solve \textbf{P3} to find a feasible solution. Furthermore, given a fixed $\m \gamma$, the equality constraint $\m M = \m Y \m \Gamma$ becomes redundant and hence, this constraint can be removed as well as the variable $\m M$ from \textbf{P3}. Moreover, when a certain variable  $\m \gamma_{i} = 0$, due to redundancy, then the corresponding row of $\m C$ and column of $\m Y$ can also be removed to obtain smaller number of variables (essentially, only consider the nonzero row of $\m C$ and column of $\m Y$). 
\end{itemize}
\vspace{-0.2cm}
The aforementioned procedures enable the SE-BnB to exploit the structure of SSP and as a result, the complexity of solving the associated SDPs required to find bounds on each node of the BnB tree can be lowered.
\vspace{-0.35cm}
\section{Some Important Extensions}\label{sec:misc}
\vspace{-0.25cm}
\subsection{Controllability and Actuator Selection}\label{ssec:actuator_controllability}
In this section, NDS controllability and  actuator selection problem (ASP) are discussed, in which we consider a similar concept to quantify controllability for a given $\gamma_l$.
\vspace{-0.10cm}
\begin{mydef}\label{def:lip_controllability}
	NDS \eqref{eq:gen_dynamic_systems} is said to be $\gamma_l$-controllable if and only if there exist $\m Q\in \mbb{S}^{n_x}_{++}$, $\m X\in \mbb{R}^{n_u\times n_x}$, and $\sigma\in\mbb{R}_{++}$ such that the LMI below is feasible
	\begin{align}
\small \hspace{-0.3cm}			\bmat{
		\m Q\m A ^{\top} + \m A\m Q - \m X ^{\top}\m B ^{\top} -\m B\m X  +\sigma \m G\m G^\top & * \\
		\m Q & -\frac{\sigma}{\gamma_l^2} \m I} & \prec 0. \label{eq:LMI_lip_con}
	\end{align}
\end{mydef}  
\vspace{-0.1cm}   
If the NDS is $\gamma_l$-controllable, then from the above LMI, the stabilizing controller gain matrix is given as $\m K = \m X\m Q^{-1}$ with control action $\m u(t)= -\m K \m x(t)$ \citep{Yadegar2018}. It is worth noting that, unlike \eqref{eq:LMI_lip_obs}, LMI \eqref{eq:LMI_lip_con} is a sufficient condition. A similar question to \textbf{Q1} now emerges: \textit{what role does $\gamma_l$ have on $\gamma_l$-controllability?} By performing a similar analysis as in Section \ref{sec:ssp} (not shown here due to space constraint), it is revealed that if $\gamma_l$ is made to be sufficiently small, then there exists at least a solution for LMI \eqref{eq:LMI_lip_con} ({if $\sigma$ is fixed}). 

Next, the relation between actuators and NDS controllability is discussed.
Let $\pi_i\in\{0,1\}$ be a binary variable representing the activation or deactivation of actuator/control node on each subsystem $i$ such that $\pi_i = 1$ if control node at $i$ is activated and $\pi_i = 0$ otherwise. For compactness, let us define $\m \pi := \left[\pi_1\,\,\pi_2\,\,\cdots\,\,\pi_N\right]^{\top}$ and $\m\Pi$ as 
$$\m \Pi := \mathrm{Blkdiag}\big(\pi_1\m I_{n_{u_1}},\pi_2\m I_{n_{u_2}},\hdots,\pi_N\m I_{n_{u_N}}\big).$$
As such, ASP for NDS \eqref{eq:gen_dynamic_systems} can be expressed as
\vspace{-0.05cm}
	\begin{align}
	\dot{\m x}(t)= \mA \m x (t) + \m G\m f(\m x) + \mB\m \Pi\m u(t),\quad\m y(t) = \mC \m x (t), \label{eq:gen_dynamic_systems_act_sel}
	\end{align}
from which a simplified high level formulation for ASP can be formulated as follows 
\vspace{-0.05cm}
\begin{align*}
(\mathbf{P5})\;\;\;	\minimize \;\;\;&  \m c^{\top} \boldsymbol \pi+ \mathrm{ControlObjective} \\
\subjectto  \;\;\;& \eqref{eq:gen_dynamic_systems_act_sel},\; \m \pi\in \mathcal{G}_\pi,\; \mathrm{ControlConstraints}. 
\end{align*}
The goals in the above are threefold: \textit{(i)} performing feedback stabilizing control on NDS \eqref{eq:gen_dynamic_systems_act_sel} while \textit{(ii)} utilizing smallest number of actuators as possible (or satisfying a given constraint over the collections of library of actuators) and \textit{(iii)} optimizing a specific control metric. In \textbf{P5}, vector $\m c\in\mbb{R}^N_{+}$ assigns weights for each actuator $\pi_i$ whereas $\mathcal{G}_\pi\subseteq \{0,1\}^N$ represents actuator's logistic constraints.
The ASP for Lipschitz NDSs can be formulated as
\begin{subequations}\label{eq:act_sel_con}
	\begin{align}
	&\hspace*{-0.3cm}(\mathbf{P6})\;	\minimize_{ \m Q, \m X, \epsilon, \m \pi}\;\;  \m c^{\top} \boldsymbol \pi\\
	&\hspace*{-0.3cm}\subjectto  \;\;	  \begin{bmatrix}
	\m Q\m A ^{\top} + \m A\m Q   +\sigma \m G\m G^\top  & \\ 
	- \m X ^{\top}\m\Pi\m B ^{\top} -\m B\m\Pi\m X & *\\
	\m Q & -\frac{\sigma}{\gamma_l^2} \m I \end{bmatrix} \preceq 0 \label{eq:act_sel_con_1} \\ 
	&\quad \quad\;\;\m Q\succ 0, \;\sigma > 0,\; \m \pi\in \mathcal{G}_\pi,\;
	\m \pi \in \{0,1\}^N.\label{eq:act_sel_con_2}\vspace*{-0.05cm}
	\end{align}
\end{subequations}
In \textbf{P6} the objective is to minimize the number of the activated actuators while \textit{(a)} finding a stabilizing controller gain matrix $\mK$ and \textit{(b)} satisfying actuator's logistic constraints. 
Notice that \textbf{P6} is also a nonconvex optimization problem with a MIBMI constraint. 
Further analysis shows that, in general, in order to employ less number of actuators, one should obtain the least possible Lipschitz constant.    

\vspace{-0.25cm}
\subsection{NDS Parameterization for Non-Lipschitz Systems}\label{ssec:NDS_parameterization}
Note that many observer/controller designs for function sets other than Lipschitz continuous such as one-sided Lipschitz, quadratically inner-bounded, bounded Jacobian, and quadratically bounded---see \citep{Phanomchoeng2010b,Abbaszadeh2010,guo2017decentralized} for notable examples---can also be considered and utilized to construct SASPs. 
For example, the nonlinearities in a NDS may satisfy one-sided Lipschitz (OSL) and quadratically inner-bounded (QIB) conditions---see Definition \ref{def:osl_qib}. 
\vspace{-0.1cm}
\begin{mydef}[OSL \& QIB]\label{def:osl_qib}
	The mapping $\m f :\mathbb{R}^{n_x}\rightarrow \mathbb{R}^{n_g}$ in \eqref{eq:gen_dynamic_systems} is locally OSL in $\mathbfcal{X}$ if for any $\m x, \hat{\m x}\in \mathbfcal{X}$ then
	\begin{subequations}\label{eq:osl_qib_def}
		\vspace{-0.1cm}
		\begin{align*}
			\langle \m G(\m f(\m x)-\m f(\hat{\m x})),\m x-\hat{\m x}\rangle\leq \gamma_s \norm{\m x - \hat{\m x}}_2^{2}, 
		\end{align*}
		for $\gamma_s\in\mbb{R}$ and QIB in $\mathbfcal{X}$ if for any $\m x, \hat{\m x}\in \mathbfcal{X}$ it holds that
		\begin{align*}
			&\langle\m G(\m f(\m x)-\m f(\hat{\m x})),\m G(\m f(\m x)-\m f(\hat{\m x}))\rangle \leq \label{eq:qib_def}\\ &\qquad\gamma_{q1} \norm{\m x - \hat{\m x}}_2^{2} +\gamma_{q2}\langle\m G(\m f(\m x)-\m f(\hat{\m x})),\m x-\hat{\m x}\rangle, \nonumber
		\end{align*}
		for $\gamma_{q1},\gamma_{q2} \in \mbb{R}$. 
	\end{subequations}	
\end{mydef}
\vspace{-0.1cm} 
In Definition \ref{def:osl_qib}, the notation $\langle\cdot,\cdot\rangle$ denotes the standard inner product. 
By considering OSL and QIB, the parameterization of NDSs entails to finding the constants $\gamma_s$, $\gamma_{q1}$, and $\gamma_{q2}$. These constants can be computed either analytically or numerically via stochastic point-based method or deterministic interval-based method. If numerical approach is pursued, then the computation of these constants reduces to solving global optimization problems. Our particular work in \cite{nugroho2020nonlinear} investigates NDSs parameterization using interval-based global maximization.

\vspace{-0.25cm}
\subsection{From OSL and QIB to Robust Selection}\label{ssec:SASP_robustness}
Now we demonstrate how the proposed methodology can be easily extend to solve SASPs for other class of nonlinearities. 
Now suppose that $\m f(\cdot)$ in NDS \eqref{eq:gen_dynamic_systems} is both OSL and QIB with constants $\gamma_s$, $\gamma_{q1}$, and $\gamma_{q2}$. Using observer design presented in \citep{zhang2012full}, the corresponding SSP can be formed as
\vspace{-0.05cm}
\begin{subequations}\label{eq:sen_sel_obs_osl_qib}
	\begin{align}
		&\hspace*{-0.2cm}\;	\minimize_{ \m P, \m Y, \epsilon_1, \epsilon_2, \m \gamma}\;\;  \m c^{\top} \boldsymbol \gamma \\
		&\hspace*{-0.2cm}\subjectto  \;\;	  \begin{bmatrix}
			\m 	A^\top \bm P + \bm P\m A + \epsilon_1\gamma_s\m I \\+\epsilon_2\gamma_{q1}\m I- \m Y\m \Gamma \m C- \m C ^{\top}\m \Gamma\m Y ^{\top} &* \\ 
			\m G^{\top}\m P+\dfrac{\gamma_{q2}\epsilon_2-\epsilon_1}{2}\m I & -\epsilon_2 \m I  
		\end{bmatrix}  \preceq 0 \label{eq:sen_sel_obs_osl_qib1} \\ 
		&\quad \quad \quad \quad\;\;\m P\succ 0, \;\epsilon_1,\epsilon_2 > 0,\; \m \gamma\in \mathcal{G}_\gamma,\;
		\m \gamma \in \{0,1\}^N,\label{eq:sen_sel_obs_osl_qib2}\vspace*{-0.05cm}
	\end{align}
\end{subequations}
while the SSP for discrete-time NDSs is formulated as \citep{Zhang2012anote}
\vspace{-0.05cm}
\begin{subequations}\label{eq:sen_sel_obs_osl_qib_discrete}
	\begin{align}
		&\hspace*{-0.2cm}\;	\minimize_{ \m P, \m Y, \epsilon_1, \epsilon_2, \m \gamma}\;\;  \m c^{\top} \boldsymbol \gamma \\
		&\hspace*{-0.0cm}\subjectto  \;\;	 \\
		&  \begin{bmatrix}
			-\m P + \epsilon_1\gamma_s\m I + \epsilon_2\gamma_{l1}\m I & * & *\\ 
			\m P \m A- \m Y \m \Gamma \m C+\dfrac{\gamma_{q2}\epsilon_2-\epsilon_1}{2}\m I & \m P-\epsilon_2\m I & *
			\\ 
			\m P \m A- \m Y \m \Gamma \m C & \m O & -\m P
		\end{bmatrix}  \preceq 0 \label{eq:sen_sel_obs_osl_qib_discrete1} \\ 
		&\;\;\m P\succ 0, \;\epsilon_1,\epsilon_2 > 0,\; \m \gamma\in \mathcal{G}_\gamma,\;
		\m \gamma \in \{0,1\}^N.\label{eq:sen_sel_obs_osl_qib_discrete2}\vspace*{-0.05cm}
	\end{align}
\end{subequations}
Notice that the problem described in \eqref{eq:sen_sel_obs_osl_qib} possesses similar nonconvexity as in \textbf{P2}. 
In addition to the above, the proposed sensor selection framework can also facilitate to include robustness in the objective. For instance,  by considering the Lipschitz condition, robust SSP with $\mathcal{L}_{\infty}$ metric can be constructed as \citep{nugroho2018journal}
\vspace{-0.05cm}
\begin{subequations}\label{eq:l_inf_theorem}
	\begin{align}
		&\minimize_{\m P, \m Y, \epsilon, \alpha, \mu_{0,1,2}} \quad \mu_0\mu_1 + \mu_2 \label{eq:l_inf_theorem_0}\\
		&\subjectto \nonumber \\
		&\bmat{ \mA^{\top}\mP + \mP\mA -\mC^{\top}\m \Gamma\mY^{\top} \\ -\mY\m \Gamma\mC+\alpha\mP+\epsilon\gamma_l^2\mI&*&*\\
			\mP & -\epsilon\mI&*\\
			\m {B_{\mathrm{w}}}^{\top}\mP-\m {D_{\mathrm{w}}}^{\top}\mY^{\top}&\mO&-\alpha\mu_0\mI} \preceq 0 \label{eq:l_inf_theorem_1}\\
		&\bmat{-\mP & * & * \\
			\mO & -\mu_2\mI & *\\
			\mZ & \mO & -\mu_1\mI}\preceq 0, \label{eq:l_inf_theorem_2}\\
		&\;\;\m P\succ 0, \;\epsilon,\mu_0,\mu_1,\mu_2 \geq 0,\;\alpha > 0,\; \m \gamma\in \mathcal{G}_\gamma,\;
		\m \gamma \in \{0,1\}^N,\label{eq:l_inf_theorem_3}\vspace*{-0.05cm}
	\end{align}
\end{subequations}
where matrices $\m {B_{\mathrm{w}}}$ and $\m {D_{\mathrm{w}}}$ represent how the disturbances are distributed, $\m Z$ is the performance matrix with $\m z(t) = \m Z \m e(t)$, and $\m e(t)$ is the estimation error.
The objective herein is to select the best sensor combination, given a fixed number of activated sensors, that minimizes the worst case disturbance attenuation $\mu = \sqrt{\mu_0\mu_1 + \mu_2}$. If the problem described in \eqref{eq:l_inf_theorem} is solved, then it is ensured that $\lim_{t\rightarrow\infty}\sup \norm{\m z(t)}_2\leq \mu \norm{\m w(t)}_{\mathcal{L}_{\infty}}$
where $\m w(t)$ denotes the disturbance vector \citep{nugroho2018journal}. 
By fixing some constants, \eqref{eq:sen_sel_obs_osl_qib} and \eqref{eq:l_inf_theorem}  share similar nonconvexity with \textbf{P2} and thus the proposed methodology can be used to solve the SSP.
Provided that the variables $\alpha$ and $\mu_0$ or $\mu_1$ are fixed, this problem along with the ones described in \eqref{eq:sen_sel_obs_osl_qib} and \eqref{eq:sen_sel_obs_osl_qib_discrete} share similar nonconvexity as \textbf{P2} and therefore, by using the proposed methodology, they can be solved using the SE-BnB algorithm.   
Next, we focus on numerically testing the SE-BnB algorithm in solving SSP for a NDS satisfying the Lipschitz condition.


\begin{figure}
	\vspace{-0.1cm}
	\centering 
	\subfloat[\label{fig:assessment_1a}]{\includegraphics[keepaspectratio=true,scale=0.76]{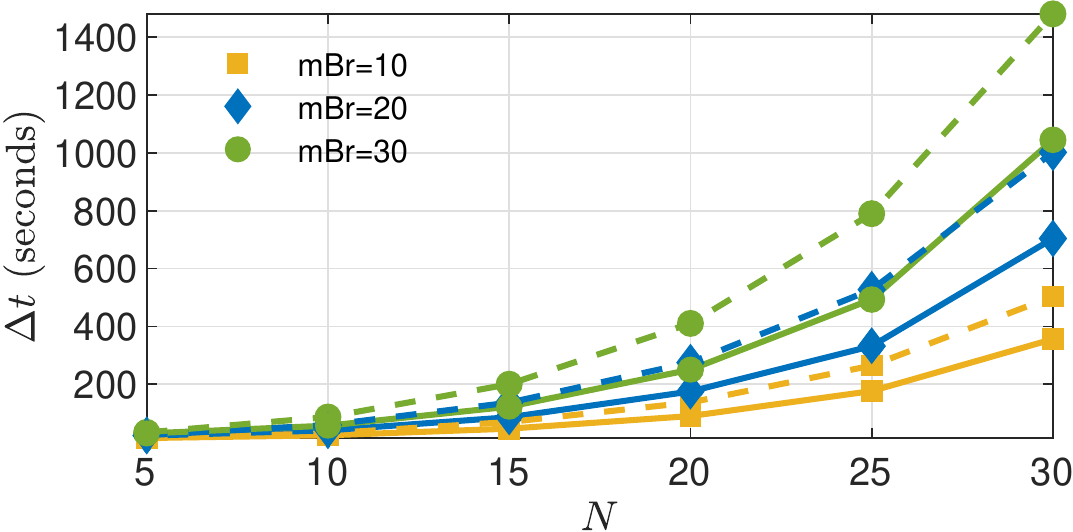}}{}\vspace{0.05cm}
	\subfloat[\label{fig:assessment_1b}]{\includegraphics[keepaspectratio=true,scale=0.76]{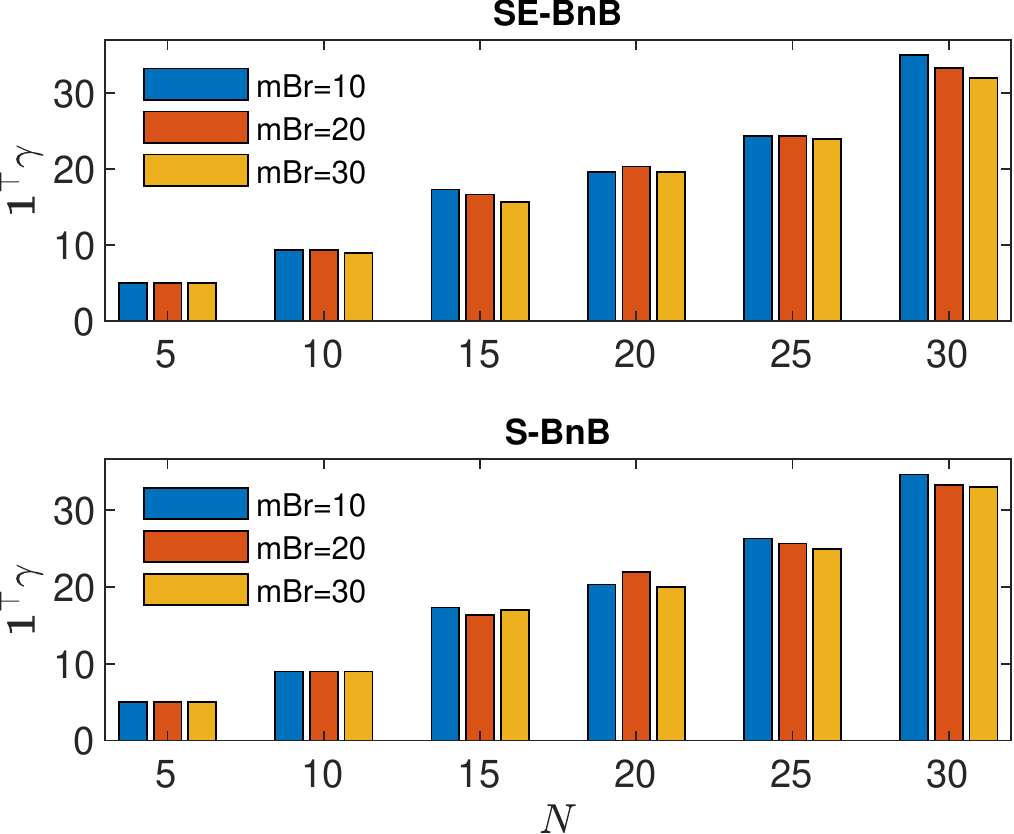}}{}\vspace{-0.1cm}\hspace{-0.0cm}\vspace{-0.2cm}
	\caption{Numerical experiment results for network of nonlinear unstable nodes to find suboptimal solutions via the SE-BnB and S-BnB: (a) computational time for different number of maximum allowable branching $\mathrm{mBr}$ and (b) the resulting sensor costs. The solid lines correspond to SE-BnB whereas dashed lines correspond to S-BnB. To compensate the heuristics on both SE-BnB and S-BnB, the numerical test is performed three times for each value of $N$ and the results shown correspond to the average values.}
	\label{fig:assessment_1}\vspace{-0.00cm}
\end{figure}

\begin{figure}
	\centering 
	\subfloat[\label{fig:assessment_4a}]{\includegraphics[keepaspectratio=true,scale=0.76]{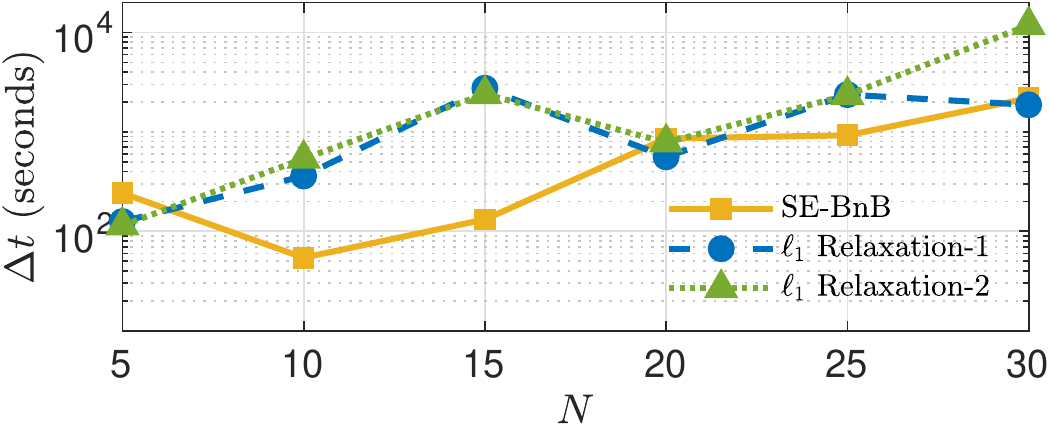}}{}\vspace{0.1cm}\vspace{-0.05cm}
	\subfloat[\label{fig:assessment_4b}]{\includegraphics[keepaspectratio=true,scale=0.76]{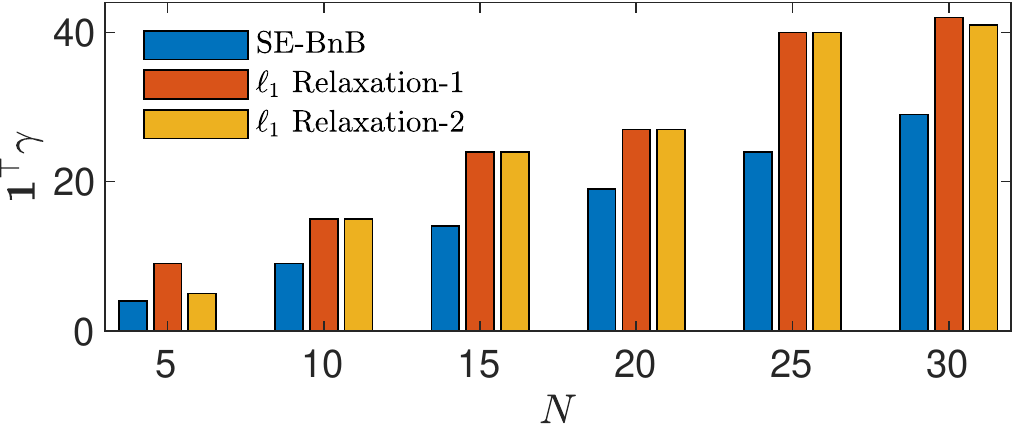}}{}\vspace{-0.2cm}\hspace{-0.0cm}\vspace{-0.00cm}
	\caption{Comparison between SE-BnB and $\ell_1$-norm relaxation to search  for suboptimal solutions: (a) computational time and (b) the resulting sensor costs. To compensate the heuristics on SE-BnB, the numerical test is performed three times and the results shown correspond to the average values.}
	\label{fig:assessment_4}\vspace{0.1cm}
\end{figure}

\vspace{-0.25cm}
\section{Numerical Assessment}\label{sec:num_test}
{In this section, we demonstrate the proposed methodology to solve SSP for a NDS on network of nonlinear unstable nodes. The simulations are performed using MATLAB R2019a, where
YALMIP's \citep{Lofberg2004} optimization package with MOSEK \citep{Andersen2000} solver are used to solve the corresponding SDPs problem.} The network of nonlinear unstable nodes considered here is adapted from \citep{Motee2008}, which is initially a network of linear systems. In order to obtain nonlinear systems, a sinusoidal function is introduced on each node such that each node $i$ has the following dynamics
\begin{align}
\hspace*{-0.2cm}\dot{\m x}_i = \bmat{\zeta_{1i} &1 \\1 & \zeta_{2i}}\m x_i + \bmat{0\\ \beta_i}\sin(x_{2i}) + \sum_{j\neq i}e^{\alpha(i,j)}\m x_j + \bmat{0\\ 1}u_i, \label{eq:nnus_dynamics}
\end{align}
where the constants $\zeta_{1i}$, $\zeta_{2i}$, and $\beta_i$ are randomly generated within $[-2,2]$, $[-2,2]$, and $[-1,1]$ respectively for each $i$. The coupling between nodes $i$ and $j$ is determined by the function of Euclidean distance $\alpha(i,j)$. This distance is randomly generated inside a box of size $5\times 5$. It can be verified that the Lipschitz constant for this system is $\gamma_l = \sqrt{N}$ where $N$ is the number of nodes. For the sake of simplicity, the weighting vector is set to be $\m c = \m 1$ where $\m Y$ is bounded such that $-10^3 \leq Y_{i,j} \leq 10^3$ for each $i,j$. From \eqref{eq:nnus_dynamics}, each node $i$ can have at most two measurements---hence for $N$ subsystems, there are at most $2N$ measurements.

\vspace{-0.25cm}
\subsection{SE-BnB and Standard BnB}
In the first case of this numerical assessment, we evaluate the performance of the SE-BnB algorithm to find suboptimal solutions for SSP. The tolerance defining optimality gap is set to be $10^{-4}$ with integer tolerance $10^{-6}$ while restricting the maximum number of allowable branching.  We also invoke a logistic constraint such that at least $20\%$ of the available measurements are equipped with sensor. We test the new SE-BnB algorithm with various number of nodes and compare the results with different values of $\mathrm{maxBranch}$ (or $\mathrm{mBr}$) against our own implementation of the standard BnB algorithm (S-BnB)---that is, only takes MISDP \textbf{P4}---while still utilizes the heuristic discussed in Section \ref{ssec:bnb_misdp}. The results of this numerical test are illustrated in Fig. \ref{fig:assessment_1}. It can be seen that, in overall, the SE-BnB requires less computational time than the S-BnB. This is due to the fact that, in the SE-BnB, the corresponding SDPs solved to obtain local lower bound and global upper bound on each BnB node have reduced complexities as opposed to the S-BnB. 
It is also showcased in Fig. \ref{fig:assessment_1b} that this reduction in computational time does not contribute to the resulting sensor costs since sensor costs are merely determined by the number of explored nodes as well as the heuristics used to find global upper bounds. Both SE-BnB and S-BnB show similar sensor costs. These costs are reduced, despite of some irregularities, as more branches are explored---these irregularities are attributed to the heuristics used in finding global upper bounds.

\vspace{-0.25cm}
\subsection{SE-BnB and $\ell_1$-Norm Relaxation}
Next, we assess the ability of our SE-BnB algorithm in finding good suboptimal solutions relative to $\ell_1$-norm relaxation, which is developed in \citep{Candes2008}, to solve a class of combinatorial optimization problems. {This technique finds its popularity in SASPs for linear systems---for example, see \citep{Argha2017}.}
This particular approach works as follows. First, note that each $\gamma_i$ determines the values of $\mathrm{col}\left(\m Y\right)_i$, i.e., $\mathrm{col}\left(\m Y\right)_i$ can be any value if and only if $\gamma_i = 1$ and zero otherwise. As such, the integer variable $\m \gamma$ can be removed from optimization variables and consequently, the SSP boils down into minimizing the number of nonzero columns of $\m Y$, i.e., $\sum_{i=1}^{n_y} \norm{\mathrm{col}\left(\m Y\right)_i}_{\ell_0}$. Since $\ell_0$-norm is nonconvex, it is replaced by  $\ell_1$-norm, yielding the following relaxed problem
\begin{subequations}\label{eq:sen_sel_obs_l1_norm}
	\vspace*{-0.2cm}
	\begin{align}
\hspace*{-0.3cm}	(\mathbf{P7})\;	\minimize_{ \m P, \m Y, \epsilon, \m \gamma}\;\; & \sum_{i=1}^{n_y}c_iw_i^{(k)}\norm{\mathrm{col}\left(\m Y\right)_i}_{\ell_1} \label{eq:sen_sel_obs_l1_norm1}\\
\hspace*{-0.3cm}	\subjectto  \;\;	 &\eqref{eq:LMI_lip_obs},	\;\m P\succ 0, \;\epsilon > 0,\label{eq:sen_sel_obs_l1_norm3}
	\vspace*{-0.25cm}
	\end{align}
\end{subequations} 
where $w_i^{(k)}$ is the corresponding weighting factor for column $i$ at iteration $k$. When $\m w^{(k)}$ is fixed, \textbf{P7} becomes a convex SDP. This weighting factor is updated at each iteration $k$ using the solution from previous iteration through the following update rule \citep{Candes2008}
\begin{align}
w_i^{(k+1)} := \frac{1}{\norm{\mathrm{col}\left(\m Y\right)_i^{(k)}}_{\ell_1} + \kappa},\; \forall i=1,2,\hdots,n_y,\label{eq:weight_update}
\end{align} 
where $\kappa > 0$ is a relatively small constant. Algorithm \ref{alg:L1-SSP}, adapted from \citep{Candes2008}, provides the detailed steps. 

In this numerical experiment, both algorithms are configured such that the objective is to find at least $N-1$ activated sensors. The $\ell_1$-norm relaxation is used with $\epsilon_w = 0.1$ and $500$ maximum iterations. Two different values for $\kappa$ are considered which correspond to two scenarios for this approach: $\ell_1$-norm relaxation-1 uses $\kappa = 10^{-4}$ while $\ell_1$-norm relaxation-2 uses $\kappa = 10^{-5}$.
The corresponding results are illustrated in Fig. \ref{fig:assessment_4}. In particular, Fig. \ref{fig:assessment_4b} indicates that our SE-BnB algorithm is able to find suboptimal sensor combinations with $N-1$ activated sensors while the $\ell_1$-norm relaxation fails to find sensor combinations under such prescribed limitation.  
On the matter of computational time, as shown in Fig. \ref{fig:assessment_4a}, the SE-BnB algorithm requires less computational time compared to the $\ell_1$-norm relaxation. {This demonstrates an advantage of the proposed SE-BnB algorithm in finding suboptimal solutions for SSP over the $\ell_1$-norm relaxation technique, at least for the network of nonlinear unstable nodes described in \eqref{eq:nnus_dynamics}.} 

\vspace{0.05cm}
\setlength{\textfloatsep}{5pt}
{\small \begin{algorithm}[t]
		\caption{\text{Iterative $\ell_1$-Norm Relaxation for SSP}}\label{alg:L1-SSP}
		\DontPrintSemicolon 
		\textbf{input:} $\epsilon_w$, $\kappa$, $\mathrm{maxIter}$\;
		\textbf{initialize:} $k = 0$, $\m w^{(1)} = \m 1$ \;
		\Do{$\abs{\norm{\m w^{(k+1)}}_{2}-\norm{\m w^{(k)}}_{2}} > \epsilon_w$ {\bf and}   $k < \mathrm{maxIter}$ \label{alg:stop}}{
			$k\leftarrow k + 1$\;
			solve \textbf{P7} in \eqref{eq:sen_sel_obs_l1_norm}, obtain $\m Y^{(k)}$\;
			\ForEach{$i\in\mbb{I}(n_y)$}{	
				update $w_i^{(k+1)}$ from $\m Y^{(k)}$ using \eqref{eq:weight_update} \;
			}
		}
		\textbf{find:} a feasible solution of LMI \eqref{eq:sen_sel_obs_l1_norm3}  where $\m Y$ corresponds to the nonzero column of $\m Y^{(k)}$ from the last iteration\;
		\textbf{output:} $\m P$, $\m Y$\;
	\end{algorithm}
}
\setlength{\floatsep}{5pt}

\subsection{Corroborating the Theory}
\begin{figure}
	\centering 
	{\includegraphics[keepaspectratio=true,scale=0.76]{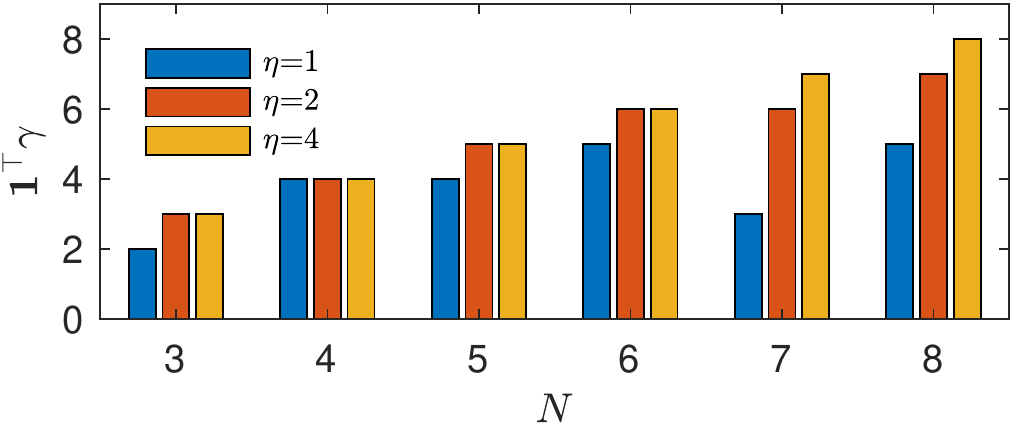}}\vspace{-0.1cm}\vspace{-0.15cm}
	\caption{Optimal sensor costs computed through the SE-BnB for different Lipschitz multiplier $\eta$ where the Lipschitz constant is given by $\gamma_l = \eta\sqrt{N}$.}
	\label{fig:assessment_3}\vspace{-0.00cm}
\end{figure}

Finally, we employ SE-BnB algorithm to find optimal solutions for SSP while using different values of Lipschitz constant. 
This numerical test is intended to find empirical relation between the value of Lipschitz constant and the number of activated sensors.
For this purpose, we consider the so called Lipschitz multiplier $\eta > 0$ such that the Lipschitz constant for $N$ number of subsystems is formulated as $\gamma_l = \eta \sqrt{N}$. Fig. \ref{fig:assessment_3} illustrates the corresponding results. It can be observed that increasing Lipschitz constant (that is, by equivalently increasing $\eta$) raises the number of activated sensors---this is indicated by the nondecreasing fashion in the number of sensor costs as the value of $\eta$ increases. This result corroborates the theoretical foundation given in Section \ref{sec:obs_ssp_sensor} and Theorem \ref{thm:lip_obs_lmi}. That is, in the context of SSP, increasing Lipschitz constant shrinks the feasible space of the SSP problem, hence requiring more sensors to stabilize the estimation error dynamics.

\vspace{-0.35cm}
\section{Paper Summary and Limitations}\label{sec:conclusion}
\textcolor{black}{
A novel general framework for dealing with SASPs for NDSs is presented. Our approach is built upon SDP formulations for observer/controller designs for various classes of NDSs developed in the literature. Specifically, this paper focuses on addressing SASPs for Lipschitz NDSs and our investigations show that smaller Lipschitz constant allows less number SAs are required to achieve stabilization in control/estimation purpose. A customized BnB algorithm, referred to as SE-BnB, which exploits problem structure and utilizing new heuristics to efficiently obtain optimal and suboptimal solutions for SASPs, is proposed. The main advantage of our framework as opposed to other approaches from the literature is its capability to obtain the optimal SAs combination by means of a BnB algorithm for stable/unstable NDS that ensures stability for either estimation and/or stabilization/tracking purposes.}

The algorithms presented in this paper come with their limitations. First, this paper is not concerned with scaling the problem for extremely large-scale systems. This is a result of using SDPs and LMIs that still have serious scalability issues---in addition to the worst case complexity of BnB routines. Second, this paper also does \textit{not} address the simultaneous selection of SAs and also assumes that all SAs are operating in ideal conditions (e.g. no saturation effect). To that end, our future work will focus on \textit{(i)} investigating different approaches to solve SASPs other than using BnB algorithm such as the generalized Benders decomposition \cite{Zhang2016} and branch-and-cut algorithm \cite{Kobayashi2020}, \textit{(ii)} extending the proposed approach for addressing the robust SASPs as well as the simultaneous selection of SAs through output feedback control and observer-based control policies, and \textit{(iii)} considering the effect of SAs saturation in SASPs.

\vspace{-0.25cm}

\section*{Acknowledgments}
This material is based upon work supported by the National Science Foundation (NSF) under Grants 1728629, 1917164, 2151571, and 2152450. We gratefully acknowledge NSF's support and the thorough comments and suggestions made by the editor and the reviewers.  


\vspace{-0.25cm}

\bibliographystyle{elsarticle-num} 
\bibliography{bib_file}

%
%

\appendix
%

\vspace{-0.1cm}
\section{Adjoint Mapping for Symmetric Matrices}\label{appdx:A}
Consider $T: \mathcal{V}\rightarrow\mathcal{W}$ where $\mathcal{V}$ and $\mathcal{W}$ are finite dimensional vector spaces with the associated inner product given as $\langle\cdot,\cdot\rangle_{\mathcal{V}}$ and $\langle\cdot,\cdot\rangle_{\mathcal{W}}$, respectively. For every $\m w\in \mathcal{W}$, the adjoint of $T$, which is denoted by $T^{\star}$, is a mapping such that $\langle T(\m v),\m w\rangle_{\mathcal{V}} = \langle \m v,T^{\star}(\m w)\rangle_{\mathcal{W}}$ for a unique $\m v\in \mathcal{V}$ that satisfies $T(\m v) = \m w$. For LMIs represented by linear mapping $\mathcal{A}:\mathcal{V}\rightarrow\mbb{S}^n$, its adjoint $\mathcal{A}^{\star}:\mbb{S}^n\rightarrow\mathcal{V}$ is therefore a mapping such that for every $\m v\in\mathcal{V}$ and $\m S\in\mbb{S}^n$, it holds that 
$\langle \mathcal{A}(\m v),\m S\rangle_{\mathcal{V}} = \langle \m v,\mathcal{A}^{\star}(\m S)\rangle_{\mbb{S}^n}$; see reference \citep{balakrishnan2003semidefinite}.

\vspace{-0.2cm}
\section{Proof of Theorem \ref{thm:lip_obs_lmi}}\label{appdx:B}
%
The proof relies on the notion of adjoint mapping for both finite-dimensional vector spaces and symmetric matrices, as discussed in \ref{appdx:A}. 
Let $\mathcal{A}: \mbb{S}^{n_x}\times \mbb{R}^{n_xn_y}\rightarrow \mbb{S}^{2n_x}\times\mbb{S}^{n_x}$ be expressed as $\mathcal{A}(\m P, \m y_v) = \mathcal{A}_{P}(\m P) + \mathcal{A}_{Y}({\m y_v})$ for $\m P\in \mbb{S}^{n_x}$ and $\m y_v\in \mbb{R}^{n_xn_y}$. First, it can be proven that for $\mathcal{A}_{P}(\m P)$ described in \eqref{eq:obs_LMI_ref_2} and matrix $\m Z$ given in \eqref{eq:lip_obs_lmi_thm_1}, $\langle \mathcal{A}_{P}(\m P),\m Z \rangle_{\mathrm{dom}(\mathcal{A})}$ is equal to 
\begin{subequations}
	\begin{align*}
	\langle \m P,-\m Z_1\m A^\top - \m A\m Z_1 - \m G\m Z_2^\top - \m Z_2\m G^\top+\m Z_4  \rangle_{\mathrm{im}(\mathcal{A})},
	\end{align*}
	yielding $\mathcal{A}_{P}^{\star}(\m Z) = -\m Z_1\m A^\top - \m A\m Z_1 - \m G\m Z_2^\top - \m Z_2\m G^\top+\m Z_4$. {Second, let us  reformulate $\mathcal{A}_{Y}({\m y_v})$ described in \eqref{eq:obs_LMI_ref_3}  into}
	\begin{align}
	\mathcal{A}_{Y}({\m y_v}) &= \sum_{i=1}^{n_xn_y} y_{v,i}\bar{\m C}_{D,i},\label{eq:lip_obs_lmi_proof_1}
	\end{align}
	where $\bar{\m C}_{D,i}$ in \eqref{eq:lip_obs_lmi_proof_1} is nothing but
	\begin{align*}
	\bar{\m C}_{D,i} = \mathrm{Blkdiag}\left(\bmat{\bar{\m C}_i & * \\ \m O & \m O},\m O\right). 
	\end{align*}
	Then, it is straightforward to show that
	\begin{align*}
\hspace*{-0.3cm}	&\langle \mathcal{A}_{Y}({\m y_v}),\m Z \rangle_{\mathrm{dom}(\mathcal{A})} \nonumber \\
\hspace*{-0.3cm}	&\;\;\;=\langle {\m y_v},\bmat{\mathrm{tr}(\bar{\m C}_{D,1}\m Z)\,\,\mathrm{tr}(\bar{\m C}_{D,2}\m Z)\,\,\cdots\,\,\mathrm{tr}(\bar{\m C}_{D,{n_xn_y}}\m Z)} \rangle_{\mathrm{im}(\mathcal{A})}\nonumber \\
\hspace*{-0.3cm}	&\;\;\;=\langle {\m y_v},\bmat{\mathrm{tr}(\bar{\m C}_{1}\m Z_1)\,\,\mathrm{tr}(\bar{\m C}_{2}\m Z_1)\,\,\cdots\,\,\mathrm{tr}(\bar{\m C}_{n_xn_y}\m Z_1)} \rangle_{\mathrm{im}(\mathcal{A})}.
	\end{align*}
	The above indicates that $\mathcal{A}_{Y}^{\star}({\m Z}) = \left\{\bmat{\mathrm{tr}(\bar{\m C}_{i}\m Z_1)}\right\}^{n_xn_y}_{i = 1}$. From these results, it can be confirmed that the adjoint mapping of $\mathcal{A}(\m P, \m y_v)$ is equivalent to $\mathcal{A}^\star(\m Z) = \left(\mathcal{A}_{P}^{\star}(\m Z),\mathcal{A}_{Y}^{\star}(\m Z)\right)$. The theorem of alternatives \citep[Theorem 1]{balakrishnan2003semidefinite} declares that whether there exist $\m P\in \mbb{S}^{n_x}$ and $\m y_v\in \mbb{R}^{n_xn_y}$ such that $\mathcal{A}(\m P, \m y_v)  + \mathcal{A}_0\succ 0$ or there exists $\m Z$ as given in \eqref{eq:lip_obs_lmi_thm_1} such that $\m Z \succeq 0$ and $\m Z \neq 0$ satisfying $\mathcal{A}^\star(\m Z) = 0$ and $\mathrm{tr}(\mathcal{A}_0\m Z) \leq 0$. Notice that $\mathcal{A}^\star(\m Z) = 0$ is indeed equivalent to
	\begin{align}
	\mathcal{A}^\star(\m Z) = 0 \Leftrightarrow 
	\begin{cases}
	\mathcal{A}_{P}^{\star}(\m Z) = 0 \\
	\mathcal{A}_{Y}^{\star}(\m Z) = 0 \label{eq:lip_obs_lmi_proof_0}
	\end{cases}.
	\end{align}
	The first right-hand side of \eqref{eq:lip_obs_lmi_proof_0} implies
	\begin{align}
	\hspace{-0.2cm}-\m Z_1\m A^\top - \m A\m Z_1 - \m G\m Z_2^\top - \m Z_2\m G^\top+\m Z_4 = 0,\label{eq:lip_obs_lmi_proof_2}
	\end{align}
	whereas the second condition implies that
	\begin{align}
	\hspace{-0.3cm}\mathrm{tr}(\bar{\m C}_{i}\m Z_1) = \sum_{j=1}^{n_x}\sum_{k=1}^{n_x}\bar{C}_{i,jk} Z_{1,jk}  = 0,\;\forall i\in\mbb{I}(n_xn_y).\label{eq:lip_obs_lmi_proof_3}
	\end{align}
	The results from \eqref{eq:lip_obs_lmi_proof_2} and \eqref{eq:lip_obs_lmi_proof_3} establish \eqref{eq:lip_obs_lmi_thm_2} and \eqref{eq:lip_obs_lmi_thm_3}, respectively. Next, note that $\mathrm{tr}(\mathcal{A}_0\m Z) \leq 0$ implies
	\begin{align}
	\mathrm{tr}(\mathcal{A}_0\m Z) = \mathrm{tr}(-\epsilon\gamma_l^2\m Z_1) + \mathrm{tr}(\epsilon\m Z_3) \leq 0.\label{eq:lip_obs_lmi_proof_4}
	\end{align}
	Since $\epsilon > 0$, then \eqref{eq:lip_obs_lmi_proof_4} establishes \eqref{eq:lip_obs_lmi_thm_4}. This concludes the proof.
\end{subequations}
\newqed

\vspace{-0.3cm}
\section{Proof of Theorem \ref{thm:lip_obs_lmi_sensor}}\label{appdx:C}
The proof of the proposition is similar to that of Theorem \ref{thm:lip_obs_lmi}. The only difference lies in the adjoint of $\mathcal{A}_{Y,\Gamma}({\m y_v})$. From \eqref{eq:obs_LMI_ref_gamma}, it is not difficult to show that
\begin{align*}
&\langle \mathcal{A}_{Y}({\m y_v}),\m Z \rangle_{\mathrm{dom}(\mathcal{A})} \nonumber \\
&\;\;\;=\langle {\m y_v},\bmat{\mathrm{tr}(\bar{\gamma}_1\bar{\m C}_{1}\m Z_1)\,\,\cdots\,\,\mathrm{tr}(\bar{\gamma}_{n_xn_y}\bar{\m C}_{n_xn_y}\m Z_1)} \rangle_{\mathrm{im}(\mathcal{A})}.
\end{align*}
Since $\bar{\gamma}_i = 0$ where $\bar{\gamma}_{i}\triangleleft{\gamma}_{j}$ for which ${\gamma}_j\notin \mathcal{S}_{\gamma}^{(a)}$, then
\begin{align*}
\mathcal{A}_{Y,\Gamma}^{\star}({\m Z}) = \left\{\bmat{\mathrm{tr}(\bar{\gamma}_i\bar{\m C}_{i}\m Z_1)}\right\}^{n_xn_y}_{i = 1},\;\forall \bar{\gamma}_{i}\triangleleft{\gamma}_{j},\;\bar{\gamma}_j\in \mathcal{S}_{\gamma}^{(a)},
\end{align*}  
which is equivalent to having $\m Z_1\in \mathcal{Z}\big(\mathcal{S}_{\gamma}^{(a)}\big)$ where $\mathcal{Z}\big(\mathcal{S}_{\gamma}^{(a)}\big)$ is defined in \eqref{eq:lip_obs_lmi_sensor}, thus completing the proof.
\newqed

\vspace{-0.3cm}
\section{Proof of Theorem \ref{thm:misdp}}\label{appdx:Z}
The equality constraint $\m M = \m Y \m \Gamma$ in \textbf{P3} is equivalent to
\begin{align}
	M_{ij} = Y_{ij}\gamma_j \Leftrightarrow 
	M_{ij} = \begin{cases}
		Y_{ij}, &\text{if}\;\;\gamma_j = 1 \\
		0, &\text{if}\;\;\gamma_j = 0, \label{eq:thm_1_proof_1}
	\end{cases}\vspace{-0.0cm}
\end{align}
where $Y_{ij}\in\left[\barbelow{Y}_{ij},\bar{Y}_{ij}\right]$ for all $i,j$ for McCormick's reformulation and $Y_{ij}\in\left[-L,L\right]$ for all $i,j$ for big-M. In what follows, we apply McCormick's relaxation to transform the consequence in \eqref{eq:thm_1_proof_1} into convex MISDP as it generalizes the big-M method.
To that end, by realizing that $\bar{Y}_{ij}-Y_{ij}$, $Y_{ij}-\barbelow{Y}_{ij}$, $1-\gamma_j$, and $\gamma_j$ are all nonnegative and noting that $M_{ij} = Y_{ij}\gamma_j$, we get 
\begin{subequations}\label{eq:thm_1_proof_2} 
	\vspace{-0.0cm}
	\begin{align}
		\left(\bar{Y}_{ij}-Y_{ij}\right)\left(1-\gamma_j\right) &\geq 0  \nonumber \\
		\Leftrightarrow M_{ij} &\geq Y_{ij}+\bar{Y}_{ij}\left(\gamma_j-1\right) \label{eq:thm_1_proof_2a} \\
		\left(\bar{Y}_{ij}-Y_{ij}\right)\gamma_j &\geq 0  
		\Leftrightarrow\bar{Y}_{ij}\gamma_j \geq M_{ij}   \label{eq:thm_1_proof_2b} \\
		\left(Y_{ij}-\barbelow{Y}_{ij}\right)\left(1-\gamma_j\right) &\geq 0  \nonumber \\
		\Leftrightarrow -M_{ij}   &\geq -Y_{ij}+\barbelow{Y}_{ij}\left(1-\gamma_j\right) \label{eq:thm_1_proof_2c} \\
		\left(Y_{ij}-\barbelow{Y}_{ij}\right)\gamma_j &\geq 0  
		\Leftrightarrow M_{ij} \geq \barbelow{Y}_{ij}\gamma_j.   \label{eq:thm_1_proof_2d} 
	\end{align}
\end{subequations}
Next, notice that there are only two possible values for $\gamma_j$. That is, for $\gamma_j=1$, then substituting $\gamma_j = 1$ to \eqref{eq:thm_1_proof_2} yields
\begin{equation*}
	\left.\begin{aligned}
		Y_{ij} \geq &\,M_{ij} \geq Y_{ij}   \\
		\bar{Y}_{ij} \geq &\,M_{ij} \geq \barbelow{Y}_{ij} 
	\end{aligned}\right\} \Rightarrow M_{ij} = Y_{ij}. \label{eq:eq:thm_1_proof_3}
\end{equation*}
On the other hand, for $\gamma_j=0$, substituting $\gamma_j = 0$ to \eqref{eq:thm_1_proof_2} yields
\begin{equation*}
	\left.\begin{aligned}
		0 \geq &\,M_{ij} \geq 0  \\
		\bar{Y}_{ij} \geq &\,Y_{ij} \geq \barbelow{Y}_{ij} 
	\end{aligned}\right\} \Rightarrow M_{ij} = 0. \label{eq:eq:thm_1_proof_4}
\end{equation*}
Since this holds for all $i,j$, then \eqref{eq:thm_1_proof_1} and \eqref{eq:thm_1_proof_2} are equivalent. For big-M method, one can obtain a similar equivalency by substituting $\barbelow{Y}_{ij} = -L$ and $\bar{Y}_{ij} = L$ in \eqref{eq:thm_1_proof_2}. 
Next, define $\m\sigma_1$, $\m\sigma_2$, $\m\Omega'$ and $\m\Omega$ as follows
\begin{align*}
	\m\sigma_1 &:= \bmat{1&-1&1&-1}^{\top},\;\\
	\m\sigma_2 &:= \bmat{-1&1&0&0}^{\top}, \\
	\m\Omega'&:= \underbrace{\bmat{\mathrm{vec}(\m \Omega)&\hdots&\mathrm{vec}(\m \Omega)}}_{n_y\, \text{times}},   \\
	\m \Omega &:= \hspace{-0.1cm}\bmat{\m \omega_{11}&\m \omega_{12}&\cdots & \m \omega_{1n_y}\\
		\m \omega_{21}&\m \omega_{22}&\cdots & \m \omega_{2n_y}\\
		\vdots&\vdots&\ddots&\vdots\\
		\m \omega_{n_x1}&\m \omega_{n_x2}&\cdots & \m \omega_{n_xn_y} }\hspace{-0.1cm},\,
\end{align*}
where $\m \omega_{ij}$ is given by
\begin{align*}
	\m \omega_{ij} := \begin{cases}
		\hspace{-0.0cm}\bmat{-\barbelow{Y}_{ij}&\bar{Y}_{ij}&
			-\bar{Y}_{ij}&\barbelow{Y}_{ij}}^\top, &\;\text{for McCormick},\\
		\hspace{-0.0cm}\bmat{L&L&
			-L&-L}^\top, &\;\text{for Big-M}.\\
	\end{cases}
\end{align*} 
By using the above notations, \eqref{eq:thm_1_proof_2} can be written as $\m \Xi \,\m \nu \leq \m\varphi$ where 
where matrix $\m \Xi\in\mbb{R}^{(2n_xn_y+n_y)\times(4n_xn_y)}$ is given as
\begin{subequations}\label{eq:xi_nu_varphi}
	\begin{align}
		\m \Xi := \bmat{\m I_{n_xn_y}\otimes \m\sigma_1\;\vline  \;\m I_{n_xn_y}\otimes \m\sigma_2\;\vline  \;\m\Omega'\odot\mI_{n_y}\otimes\mI_{4n_x}},
	\end{align}
	vector $\m \nu\in\mbb{R}^{2n_xn_y+n_y}$ is given as 
	\begin{align}
		\m \nu := \bmat{\mathrm{vec}(\m M)^{\top}\;\;\mathrm{vec}(\m Y)^{\top}\;\;\m \gamma^{\top}}^{\top}, 
	\end{align}
	and vector $\varphi\in\mbb{R}^{4n_xn_y}$ is given as
	\begin{align}
		\m\varphi := \mathrm{vec}(\m \Omega)\odot\m I_{n_xn_y}\otimes\bmat{1&1&0&0}^{\top}. 
	\end{align}
\end{subequations} 
This yields the following problem
\begin{subequations}\label{eq:sen_sel_obs_misdp}
	\begin{align}
		\;	\minimize_{ \m P, \m Y, \m M, \epsilon, \m \gamma}\;\; &\m c^{\top} \boldsymbol \gamma \label{eq:sen_sel_obs_misdp_0}\\[0.001\baselineskip]
		\subjectto \;\; &\eqref{eq:sen_sel_obs_const_1},\,\eqref{eq:sen_sel_obs_2},\,\m Y \in B(\m Y),\,\m \Xi \,\m \nu \leq \m \varphi,\label{eq:sen_sel_obs_misdp_1} 
	\end{align}
\end{subequations}  
Since other constraints except $\m M = \m Y\m \Gamma$ in \textbf{P3} are convex, then \textbf{P3} and the problem described in \eqref{eq:sen_sel_obs_misdp} are equivalent.
The next step is to transform problem \eqref{eq:sen_sel_obs_misdp} into standard inequality form. 
To do so, notice that the logistic constraint in \eqref{eq:sen_sel_obs_2} can be implicitly represented by $\m H^\top \m \gamma \leq \m h$ in which $\m H\in\mbb{R}^{n_y\times n_h}$ and $\m h\in\mbb{R}^{n_h}$ whereas $\m Y \in B(\m Y)$ can be represented by $\m \Phi ^\top \mathrm{vec}(\m Y) \leq \m \psi$ where $\m \Phi = \bmat{\m I & -\m I}\in \mbb{R}^{n_xn_y\times 2n_xn_y}$ and $\m\psi = \bmat{\mathrm{vec}\left(\bar{\m Y}\right)^\top&\mathrm{vec}\left(\barbelow{\m Y}\right)^\top}$ for McCormick's relaxation---similarly for big-M.
Next, define $\mbb{I}_v(\m P)$, $\mbb{I}_v(\epsilon)$, $\mbb{I}_v(\m M)$, $\mbb{I}_v(\m Y)$, and $\mbb{I}_v(\m \gamma)$ as the set of indices for $\m v$---given in \eqref{eq:vector_v}---that correspond to $\m P$, $\epsilon$, $\m M$, $\m Y$, and $\m \gamma$, which can be combined into index set $\mbb{I}(\m v)$. Next, define $$\mathcal{E}_P := \{\m E_{P,1}, \m E_{P,2},\hdots,\m E_{P,\bar{n}_x}\}$$ as the set of standard basis in $\mbb{S}^{n_x}$. Likewise, $$\mathcal{E}_M := \{\m E_{M,1}, \m E_{M,2},\hdots,\m E_{M,n_xn_y}\}$$ is defined as the set of standard basis in $\mbb{R}^{n_x\times n_y}$. As such, the objective function \eqref{eq:sen_sel_obs_misdp_0} can be written as $\m c^{\top}_v \m v$ where vector $\m c_v$ is constructed as $\bmat{\m 0^{1\times(\bar{n}_x+1+2n_xn_y)}\;\;\m c^{\top}}^{\top}$ while the linear terms in problem \eqref{eq:sen_sel_obs_misdp} can be written as
\begin{align*}
	\mathcal{L}_{v}(\m v) := \sum_{i\in \mbb{I}(\m v)} v_i\,\mathrm{Blkdiag}\left(\m A^{(1)}_i,\m A^{(2)}_i,\m A^{(3)}_i,\m A^{(4)}_i\right),
\end{align*} 
where $\m A^{(1)}_i$, $\m A^{(2)}_i$, $\m A^{(3)}_i$, $\m A^{(4)}_i$ are detailed as follows
\begin{align*}
	\m A^{(1)}_i &=
	\begin{cases}
		\mathrm{Blkdiag}\left(\begin{bmatrix}
			-\m A ^{\top}\m E_{P,i} - \m E_{P,i}\m A   & *\\
			-\m G ^{\top}\m E_{P,i} & \m O \end{bmatrix},\m E_{P,i},\m O\right),  \\
		\qquad\qquad\qquad\qquad\qquad\qquad\qquad\;\;\,\;\text{if}\; i\in \mbb{I}_v(\m P) &\\
		\mathrm{Blkdiag}\left(-\gamma_{l}^2\m I,\m I,\m O,\m I\right),\qquad\qquad\,\text{if}\; i\in \mbb{I}_v(\epsilon)& \\
		\mathrm{Blkdiag}\left(\m E_{M,i}\m C + \m C^\top\m E_{M,i}^\top,\m O\right),\;\;\text{if}\; i\in \mbb{I}_v(\m M)& \\
		\m O, \qquad\qquad\qquad\qquad\qquad\qquad\quad\;\;\text{otherwise}
	\end{cases}\hspace{-0.4cm}, \\
	\m A^{(2)}_i &=
	\begin{cases}
		\mathrm{Diag}\left(-H_{1i},-H_{2i},\hdots,-H_{n_hi}\right), &\,\text{if}\;i\in \mbb{I}_v(\m Y) \\
		\m O, &\,\text{otherwise}
	\end{cases}, \\
	\m A^{(3)}_i &=
	\begin{cases}
		\mathrm{Diag}\left(-\Phi_{1i},-\Phi_{2i},\hdots,-\Phi_{2n_xn_yi}\right), &\hspace{-0.2cm}\text{if}\;i\in \mbb{I}_v(\m Y) \\
		\m O, &\hspace{-0.2cm}\text{otherwise}
	\end{cases}\hspace{-0.1cm},\\
	\m A^{(4)}_i&=
	\begin{cases}
		\mathrm{Diag}\left(-\Xi_{1i},-\Xi_{2i},\hdots,-\Xi_{4n_xn_yi}\right), &\hspace{-0.2cm}\text{if}\;i\in \mbb{I}_v(\m z) \\
		\m O, &\hspace{-0.2cm}\text{otherwise}
	\end{cases}\hspace{-0.1cm},
\end{align*}
where $\mbb{I}_v(\m z) := \left(\mbb{I}_v(\m M),\mbb{I}_v(\m Y),\mbb{I}_v(\m \gamma)\right)$. Accordingly, the constant terms in problem \eqref{eq:sen_sel_obs_misdp} can be written as $$\mathcal{C}_v := \mathrm{Blkdiag}\left(\m C^{(1)},\m C^{(2)},\m C^{(3)},\m C^{(4)}\right)$$ where $\m C^{(1)}$, $\m C^{(2)}$ $\m C^{(3)}$, $\m C^{(4)}$ are constructed as
\begin{align*}
	\m C^{(1)} &= -\kappa\m I,\quad 
	\m C^{(2)} = \mathrm{Diag}\left(h_{1},h_{2},\hdots,h_{n_h}\right),\\
	\m C^{(3)} &= \mathrm{Diag}\left(\psi_{1},\psi_{2},\hdots,\psi_{2n_xn_y}\right),\;\; \\
	\m C^{(4)} &= \mathrm{Diag}\left(\varphi_{1},\varphi_{2},\hdots,\varphi_{4n_xn_y}\right),
\end{align*}
in which $\kappa > 0$ is a relatively small constant to ensure positive definiteness. The resulting problem can now be presented as
\begin{align*}
	\minimize_{\m v}\;\; & \m c^{\top}_v \m v \\
	\subjectto  \;\;	 & \mathcal{L}_v(\m v) + \mathcal{C}_v\succeq 0 , \\
	\;\;& v_i\in \{0,1\},\;\;\forall i\in\mbb{I}_v(\m \gamma),
\end{align*}
which is \textbf{P4}. This ends the proof.
\newqed

\end{document}